\documentclass[twocolumn]{emulateapj}

\usepackage{natbib,amsmath}
\usepackage{color}
\usepackage{multirow}

\submitted{Accepted for Publication in The Astrophysical Journal, Feb 17, 2015}

\begin{document}

\newcommand{\tnm}{\tablenotemark}
\newcommand{\tnt}{\tablenotetext}
\newcommand{\fnm}{\footnotemark}

\newcommand{\jcap}{Journal of Cosmology and Astroparticle Physics}
\newcommand{\nar}{New Astronomy Review}
\newcommand{\pasa}{Publications of the Astron. Soc. of Australia}

\newcommand{\NoOfATCAGMRTPairs}{28}                              
\newcommand{\AlphaSearchRadius}{$6^{\prime}$}                    
\newcommand{\AlphaMedian}{$\alpha^{2.1}_{610\,\rm median}=0.56$} 
\newcommand{\AlphaInnerQuartile}{0.37--0.69}                     

\newcommand{\NoOfAtcaSources}{1934}                         
\newcommand{\MaximumRadius}{$r_{\rm max}=10^{\prime}$}      
\newcommand{\BinLow}{$28\,\rm\mu Jy$}                       
\newcommand{\BinHigh}{$1.65\,\rm Jy$}

\newcommand{\NoOfRadioInRFH}{346}                                                            
\newcommand{\AtcaContamRFH}{$ 39 ^ { +6 }_{ -6 }\,\rm mJy$}                                  
\newcommand{\AtcaContamArOneRFH}{2.1--$8.1\,\rm mJy$} 
\newcommand{\NoOfSMGInRFH}{36}                                                               
\newcommand{\LabocaContamRFH}{ $ 46 ^{+ 24 }_{ -32 }\,\rm mJy$}                              

\newcommand{\TotalSmgSolidAngle}{$0.10\,\rm deg^{2}$}                                        
\newcommand{\SmgCounts}{$N(>10\,{\rm mJy})=223 \pm 49 \,\rm deg^{-2}$}                       
  
\newcommand{\SmgContamLaboca}{$26_{-26}^{+21}\,\rm mJy$}   
\newcommand{\SmgContamArOne}{$2\pm 2\,\rm mJy$}            
\newcommand{\SmgContamArTwo}{$6\pm 5\,\rm mJy$}            
\newcommand{\SmgContamLabocaFrac}{$18\%$}                   
\newcommand{\SmgContamArOneFrac}{$2\%$}                    

\newcommand{\RadioContamRadio}{$28\pm 8\,\rm mJy$}   
\newcommand{\RadioContamArOne}{$3\pm 1\,\rm mJy$}            
\newcommand{\RadioContamArTwo}{$2\pm 2\,\rm mJy$}            
\newcommand{\RadioContamArOneFrac}{$ 3\% $}                

\newcommand{\LaflaccaRedshift}{$z=4.09$}
\newcommand{\LaflaccaLIR}{$L_{\rm IR}=3.19\times 10^{12}\,L_{\odot}$}

\newcommand{\RawEfficiency}{$0.50$}
\newcommand{\IterativeEfficiency}{$0.97$}
\newcommand{\RawSZEEfficiency}{$0.10$}
\newcommand{\IterativeSZEEfficiency}{$0.54$}
\newcommand{\FourierSZEEfficiency}{$0.53$}

\newcommand{\MeanVpec}{$\left<v_{p}\right>=153 \pm 383\,\rm km\,s^{-1}$}
\newcommand{\vpecVariance}{$1449\,\rm km\,s^{-1}$}
\newcommand{\fracMassError}{$\left<\Delta M / M\right> = 3.5 \pm 0.8\%$}

\title{The Atacama Cosmology Telescope: \\ 
The LABOCA/ACT Survey of Clusters at All Redshifts}

\author{
Robert~R.~Lindner\altaffilmark{1,2}\altaffilmark{$\dagger$},
Paula~Aguirre\altaffilmark{3},
Andrew~J.~Baker\altaffilmark{1},
J.~Richard~Bond\altaffilmark{4},
Devin~Crichton\altaffilmark{5},
Mark~J.~Devlin\altaffilmark{6},
Thomas~Essinger-Hileman\altaffilmark{5},
Patricio~Gallardo\altaffilmark{7,8},
Megan~B.~Gralla\altaffilmark{5,9},
Matt~Hilton\altaffilmark{10,11},
Adam~D.~Hincks\altaffilmark{12},
Kevin~M.~Huffenberger\altaffilmark{13},
John~P.~Hughes\altaffilmark{1},
Leopoldo~Infante\altaffilmark{10},
Marcos~Lima\altaffilmark{14},
Tobias~A.~Marriage\altaffilmark{5},
Felipe~Menanteau\altaffilmark{15,16},
Michael~D.~Niemack\altaffilmark{7},
Lyman~A.~Page\altaffilmark{17},
Benjamin~L.~Schmitt\altaffilmark{6},
Neelima~Sehgal\altaffilmark{18},
J.~L.~Sievers\altaffilmark{10,19},
Crist\'obal~Sif\'on\altaffilmark{20},
Suzanne~T.~Staggs\altaffilmark{17},
Daniel~Swetz\altaffilmark{21},
Axel~Wei{\ss}\altaffilmark{22}, \&
Edward J. Wollack\altaffilmark{23}                            
}

\altaffiltext{1}{Department of Physics and Astronomy, Rutgers, The State University of New Jersey, 136 Frelinghuysen Road, Piscataway, NJ 08854-8019, U.S.A.}
\altaffiltext{2}{Department of Astronomy, The University of Wisconsin-Madison, 475 N. Charter Street, Madison, WI 53706-1582, U.S.A.}
\altaffiltext{3}{School of Engineering, Pontificia Universidad Cat\'olica de Chile,  Av. Vicuna Mackenna 4068, Macul, Santiago, Chile}
\altaffiltext{4}{Canadian Institute for Theoretical Astrophysics, University of Toronto, Toronto, ON M5S 3H8, Canada}
\altaffiltext{5}{Department of Physics and Astronomy, The Johns Hopkins University, 3400 N. Charles St., Baltimore, MD 21218-2686, U.S.A.}
\altaffiltext{6}{Department of Physics and Astronomy, University of Pennsylvania, 209 South 33rd Street, Philadelphia, PA 19104, U.S.A.}
\altaffiltext{7}{Department of Physics, Cornell University, Ithaca, NY 14853, U.S.A.}
\altaffiltext{8}{Departamento de Astronom\'ia y Astrof\'isica, Facultad de F\'isica, Pontific\'ia Universidad Cat\'olica, Casilla 306, Santiago 22, Chile}
\altaffiltext{9}{Harvard-Smithsonian Center for Astrophysics, 60 Garden Street, Cambridge, MA 02138, U.S.A.}
\altaffiltext{10}{Astrophysics and Cosmology Research Unit, School of Mathematics,Statistics \& Computer Science, University of KwaZulu-Natal, Durban 4041, South Africa}
\altaffiltext{11}{Centre for Astronomy \& Particle Theory, School of Physics \& Astronomy, University of Nottingham, Nottingham NG7 2RD, UK}
\altaffiltext{12}{Department of Physics and Astronomy, University of British Columbia, 6224 Agricultural Rd., Vancouver BC V6T 1Z1, Canada}
\altaffiltext{13}{Department of Physics, Florida State University, 609 Keen Physics Building, Tallahassee, FL 32306, U.S.A.}
\altaffiltext{14}{Departamento de F\'isica Matem\'atica, Instituto de F\'isica, Universidade de S\~ao Paulo, S\~ao Paulo SP, Brazil}
\altaffiltext{15}{National Center for Supercomputing Applications, University of Illinois at Urbana-Champaign, 1205 W. Clark St., Urbana, IL 61801, U.S.A.}
\altaffiltext{16}{Department of Astronomy, University of Illinois at Urbana-Champaign, W. Green Street, Urbana, IL 61801, U.S.A.}
\altaffiltext{17}{Joseph Henry Laboratories of Physics, Jadwin Hall, Princeton University, Princeton, NJ 08544, U.S.A.}
\altaffiltext{18}{Department of Physics and Astronomy, Stony Brook, NY 11794-3800, U.S.A.}
\altaffiltext{19}{National Institute for Theoretical Physics (NITheP), University of KwaZulu-Natal, Private Bag X54001, Durban 4000, South Africa}
\altaffiltext{20}{Leiden Observatory, Leiden University, PO Box 9513, NL-2300 RA Leiden, Netherlands}
\altaffiltext{21}{NIST Quantum Devices Group, 325 Broadway Mailcode 817.03, Boulder, CO 80305, U.S.A.}
\altaffiltext{22}{Max-Planck-Institut f\"ur Radioastronomie, Auf dem H\"ugel 69, D-53121 Bonn, Germany}
\altaffiltext{23}{NASA/Goddard Space Flight Center, Greenbelt, MD 20771, U.S.A}
\altaffiltext{24}{{\it Herschel} is an ESA space observatory with science instruments provided by European-led Principal Investigator consortia and with important participation from NASA.}
\altaffiltext{$\dagger$}{rlindner@astro.wisc.edu}

\begin{abstract}
We present a multi-wavelength analysis of eleven Sunyaev Zel'dovich 
effect (SZE)-selected galaxy clusters (ten with new data) from the
Atacama Cosmology Telescope (ACT) southern survey. 
We have obtained new imaging from the Large APEX Bolometer 
Camera (345\,GHz; LABOCA) on the Atacama Pathfinder EXperiment 
(APEX) telescope, the Australia Telescope Compact Array 
(2.1\,GHz; ATCA), and the Spectral and Photometric Imaging 
Receiver (250, 350, and $500\,\rm \mu m$; SPIRE) on the 
{\em Herschel Space Observatory}\altaffilmark{23}.
Spatially resolved $345\,\rm GHz$ SZE increments with 
integrated $\rm S/N > 5$ are found in six clusters.  We 
compute $2.1\,\rm GHz$ number counts as a 
function of cluster-centric radius and find significant 
enhancements in the counts of bright sources at projected radii 
$\theta < \theta_{2500c}$.  By extrapolating in frequency, 
we predict that the combined signals from $2.1\,\rm GHz$-selected 
radio sources and $345\,\rm GHz$-selected SMGs contaminate the 
$148\,\rm GHz$ SZE decrement signal by $\sim 5\%$
and the $345\,\rm GHz$ SZE increment by $\sim 18\%$.
After removing radio source and SMG emission from the SZE 
signals, we use ACT, LABOCA, and (in some cases) new 
{\em Herschel} SPIRE imaging to place constraints on the 
clusters' peculiar velocities.  The sample's
average peculiar velocity relative to the cosmic microwave 
background is \MeanVpec{}.
\end{abstract}

\section{Introduction}
\setcounter{footnote}{0}

Galaxy clusters produce a spectral distortion in the cosmic 
microwave background (CMB) known as the Sunyaev Zel'dovich effect 
\citep[SZE; ][]{zeld69,suny70}.  The thermal Sunyaev Zel'dovich 
effect (tSZ) signal is quantified by  $Y_{\rm tSZ}\equiv \int y\,d\Omega$ 
in terms of the Compton parameter 
\begin{equation}
    \label{e-compton}
    y = \int \sigma_T\,n_e \frac{k_{\rm B}T_e}{m_ec^2}\,dl,
\end{equation}
the Thomson cross section $\sigma_T$, Boltzmann's constant
$k_{\rm B}$, the electron density $n_e$, the electron 
temperature $T_{e}$, and the line-of-sight path length $dl$.
The Compton parameter is insensitive to cluster redshift, allowing 
for the unbiased detection of massive clusters out
to large distances.  This selection is
complementary to that of surveys using optical 
richness or X-ray flux, which generally yield lower-redshift
cluster samples
\citep[for a review of the SZE in clusters, see e.g., ][]{carl02}. 
The mass function of SZE-selected galaxy clusters 
has been used to constrain the 
properties of dark energy, as well as the mean 
matter density $\Omega_m$ and amplitude of 
fluctuations $\sigma_8$ 
\citep[e.g., ][]{sehg11, bens13, hasselfield2013, plan13xx}.
 
The number of known SZE-selected clusters
is rising rapidly.  The first surveys produced samples of $\sim 20$
blind SZE-detected clusters \citep{vand10, mena10, marr11}.
A sample of 68 SZE-selected 
clusters from the equatorial footprint of the
Atacama Cosmology Telescope (ACT) 
survey has recently been presented by \citet{mena13} and 
\citet{hasselfield2013}; 
158 were detected in the first $720\,\rm deg^{2}$ of the 
South Pole Telescope (SPT) SZ survey \citep{reichardt2013}; 
677 were detected the full $2500\,\rm deg^{2}$ SPT-SZ 
survey \citep{bleem2014}; and $\sim 1200$ more all-sky 
SZE cluster candidates have been catalogued by 
the \cite{plan13xxix}.
Yet with samples ranging from
nine  \citep{sehg11} to eighteen 
\citep{vand10,bens13} clusters,
the first analyses found that
the statistical errors
on $w$ and $\sigma_8$ are already smaller
than systematic errors 
due to uncertainties in the 
$Y_{\rm SZ}$-to-mass scaling relation.
\citet{hasselfield2013} confirm this limitation by
finding that the improvements in cosmological parameters
from their sample of 68 SZE-selected clusters is mainly due
to the inclusion of dynamical mass information, not
to an increased sample size. 
Therefore, to further improve constraints from 
SZE-cluster cosmology using these larger samples, 
we need a better understanding of the
scaling between integrated SZE signal and 
cluster mass in individual systems. For example, 
\citet{sifo13} have measured the dynamical masses 
of 16 massive SZE-selected clusters (nine of which 
are in our sample) and find that disturbed or 
merging systems could be biasing the 
derived $Y_{\rm SZ}$-to-mass scaling relations.

A number of physical processes are known to
cause deviations from equilibrium scaling relations.
Cluster mergers are predicted to cause departures 
from hydrostatic equilibrium, and can produce 
transient pressure enhancements that boost the 
SZE signal \citep[e.g., ][]{poole2006,poole2007,wik08}.
Such pressure enhancements 
are invisible in X-ray observations
of the massive cluster RXJ1347-1145, for example, but 
were revealed through high-resolution SZE-imaging
to contribute $\sim 10\%$ of the bulk
signal \citep{koma01,kita04,maso10}.
Kinetic Sunyaev Zel'dovich (kSZ) signals 
due to clusters' peculiar velocities
can also introduce additional scatter to 
$Y_{\rm SZ}$ measurements.
\citet{hand12} recently achieved a  
statistical detection of the pairwise kSZ
signal from an ACT sample, but the
peculiar velocity distribution of clusters
remains unconstrained.
\citet{mroc12} found some evidence for
large kSZ distortions in the
triple merger system
MACS\,J0717.5+3745, perhaps indicating that 
high-velocity substructures in merging clusters introduce
additional deviations to $Y_{\rm SZ}$.
\citet{ruan13} predict these deviations
could bias $Y_{\rm SZ}$ results by $\sim 10\%$.
The emission from bright galaxies near the clusters
(in projection) can further bias the SZE signal.
Synchrotron emission from
star-forming galaxies and active galactic nuclei
may ``fill in'' SZE decrements.
\citet{rees12} estimate
contamination from synchrotron sources to be
$\lesssim 20\%$ in the $90\,\rm GHz$ SZE decrement 
based on high-resolution imaging
of two ACT clusters, and \citet{sayers2013b} estimate
that radio sources contaminate the $140\,\rm GHz$ SZE
decrement by $\sim 1$-$20\%$.
SZE increments, on the other hand, can be
artificially enhanced by the strong 
infrared emission from
dusty, high-redshift star-forming submillimeter galaxies
\citep[SMGs; for a review, see ][]{blai02}.
Gravitational lensing by the
clusters' potentials may increase contamination
by lensed SMGs \citep[e.g., ][]{knud08,joha11,jain2011} 
or introduce deficits in surface brightness at the
location of the cluster \citep{zemc13}, and disturbed or
merging systems may be more susceptible to lensing effects
due to their greater lensing efficiency 
\citep[e.g., ][]{meneghetti2007,zitrin2013b}.  \citet{bens03} conclude that
direct measurements of cluster peculiar velocities
in maps with angular resolution $\gtrsim 1\,\rm arcmin$ will
be limited by SMG contamination.

The number counts of the most massive 
clusters have the potential to constrain 
cosmological parameters.  
Unfortunately, these systems also 
tend to be most affected by the effects 
described above.  Massive systems will produce
the greatest gravitational lensing shear, and
in our hierarchical universe, 
they are also commonly 
disrupted by recent merging activity.  
In this work, we aim to better understand
how these considerations affect the observed SZE
signals using high-resolution submillimeter and
radio imaging of a representative sample of SZE-selected
clusters.We present new observations at 
$345\,\rm GHz$ ($19.2^{\prime\prime}$ resolution) 
with the Large APEX Bolometer 
Camera \citep[LABOCA; ][]{siri09}
on the Atacama Pathfinder EXperiment 
\citep[APEX; ][]{gusten2006} 
telescope\footnote{This publication is based on data 
acquired with the Atacama 
Pathfinder Experiment (APEX). APEX is a collaboration 
between the Max-Planck-Institut f\"{u}r Radioastronomie, the 
European Southern Observatory, and the 
Onsala Space Observatory.} 
and at $2.1\,\rm GHz$ ($5^{\prime\prime}$ resolution) with
the Australia Telescope Compact Array (ATCA) of a sample
of massive SZE-selected galaxy clusters.
We call the project ``LASCAR,'' the LABOCA/ACT Survey of Clusters at 
All Redshifts, in honor of the Lascar 
volcano near the ACT site in northern Chile.
We use these data to measure the properties of the
clusters' spatially resolved SZE increment signals, 
and quantify the degree of 
background and foreground radio and infrared 
galaxy contamination.  Section \ref{s-sample}
describes our cluster sample
Section \ref{s-obs-lascar}
presents observations and data reduction techniques. 
Section \ref{s-contam} assesses the SZE contamination by
point sources. Section \ref{s-vpec} uses the 
point-source subtracted multi-wavelength SZE maps
to place constraints on cluster peculiar velocities. 
In Section \ref{s-discuss} we discuss our results in the
context of previous work, and in Section 
\ref{s-conclusions}, we conclude. In our calculations, 
we assume a flat $\Lambda$CDM cosmology with 
$H_0=70\,\rm km\,s^{-1}\,Mpc^{-1}$, $\Omega_M=0.27$, and 
$\Omega_\Lambda=0.73$ \citep{koma11}.  

\section{Cluster sample}
\label{s-sample}

Our sample consists of eleven clusters with 
$148\,\rm GHz$ decrement signal-to-noise 
(S/N) $>4.7$ from the Atacama Cosmology Telescope 
\citep[ACT; ][]{fowl07,swetz2011} southern survey 
\citep{mena10,marr11}. 
Our sample is selected from the 15 highest-S/N ACT southern clusters, 
and includes 9 of 10 clusters not known 
before ACT or SPT; it only excludes one
cluster with $z<0.15$ and three clusters
that had been previously mapped with AzTEC 
(D. Hughes, personal communication). 
The properties are listed in Table \ref{t-clusters}.  
The systems span a
large redshift range, $z=0.3$--$1.1$,
and have masses
$M_{500c}\geq 3\times 10^{14}\,M_{\odot}$
\citep{mena10,sifo13}, where $M_{500c}=500\,(4\pi/3)\,\rho_c\,r_{500c}^3$,
and $r_{500c}$ is the radius enclosing a mass density equal to
$500\times$ the critical density of the Universe at the redshift 
of the cluster.  We define the angular radius
$\theta_{500c} = r_{500c}/D_{A}$, and use the
typical scaling relation $M_{200c} \sim 1.6\,M_{500c}$ \citep{duffy2008}
to convert from the dynamical masses of \citet{sifo13}.
Each SZE detection has been confirmed 
to be a rich optical cluster through
followup imaging by \citet{mena10}.
Included in the sample are the notable cluster 
mergers ACT\,J0102-4915, also known as ``El Gordo''
\citep{mena12}, 
and 1E0657-56 (ACT\,J0658-5557), 
the original ``Bullet'' cluster \citep{mark02}.

\section{Observations and data reduction}
\label{s-obs-lascar}

\subsection{345\,GHz APEX/LABOCA}

We obtained new LABOCA $345\,\rm GHz$ 
imaging of ten clusters in 2010--2011 
(see Table \ref{t-clusters}).
The LABOCA data for an eleventh cluster that is also detected
by ACT (ACT-CL\,J0658$-$5557)
are from the European Southern 
Observatory (ESO) archive.  The following subsections
describe the algorithms used to reduce the
LABOCA data and extract SZE signals for our full
sample of eleven clusters.

\subsubsection{LABOCA observations}

\begin{figure*}
\centering
\begin{tabular}{cc}
  \includegraphics[scale=0.45]{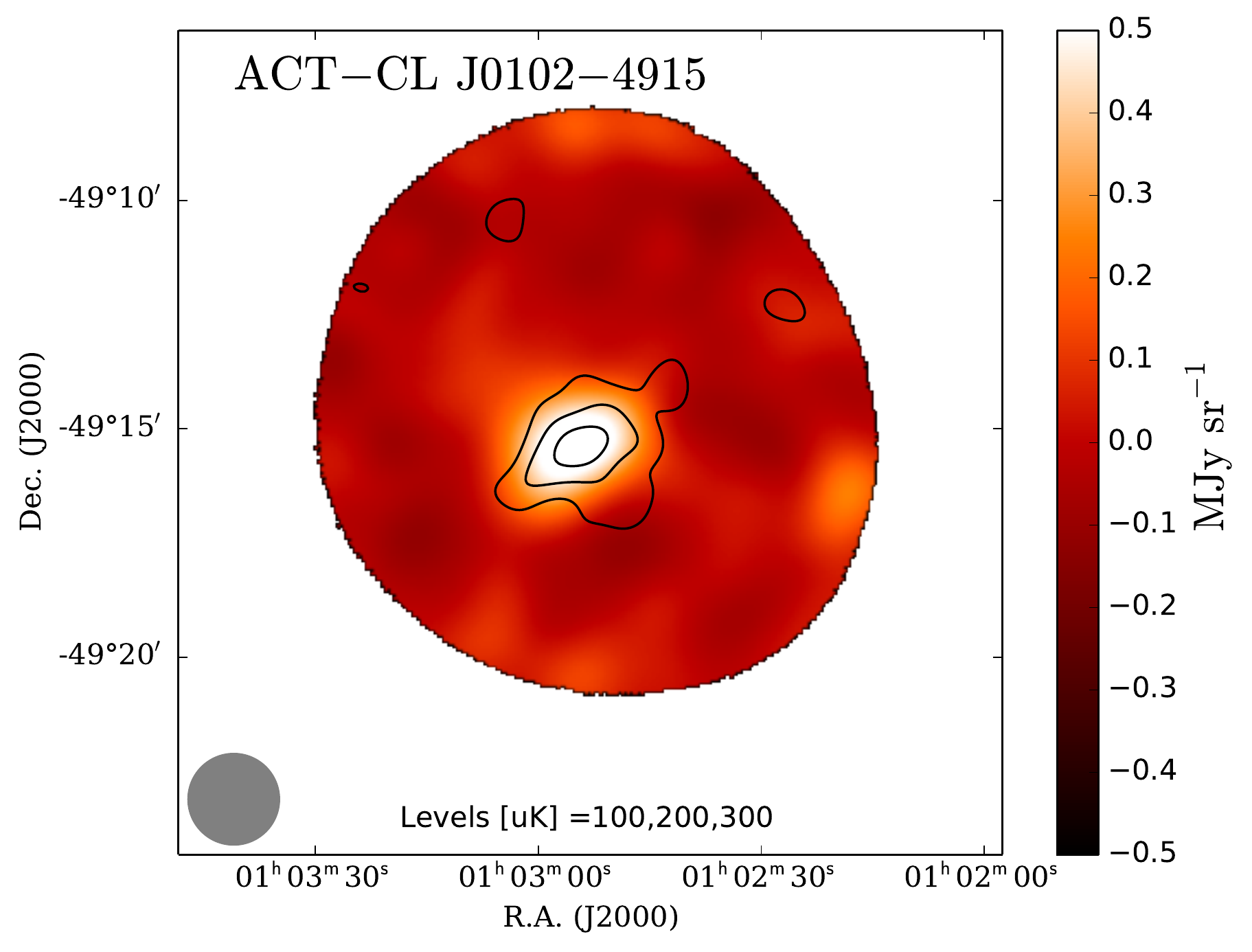} &
  \includegraphics[scale=0.45]{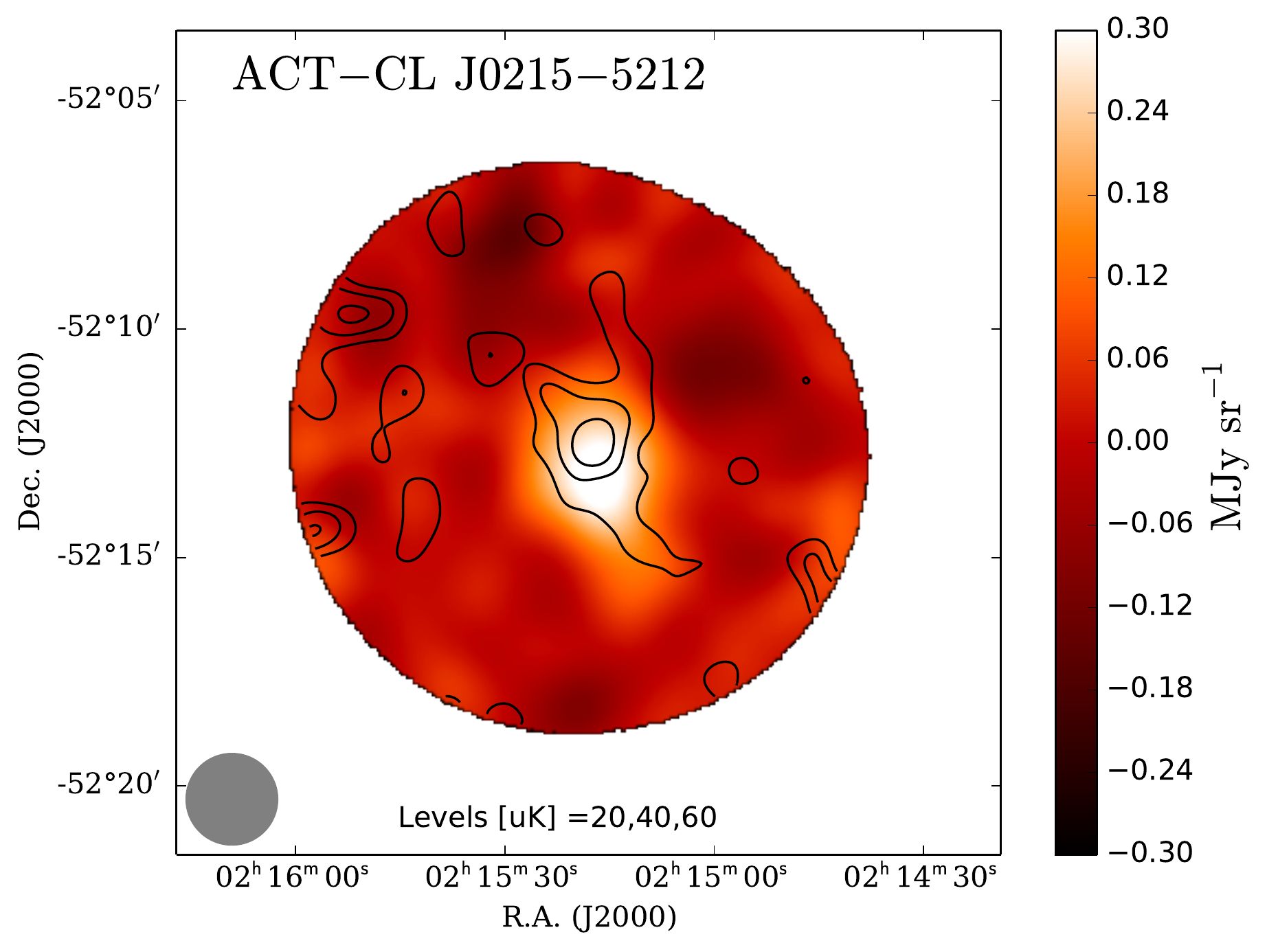} \\
  \includegraphics[scale=0.45]{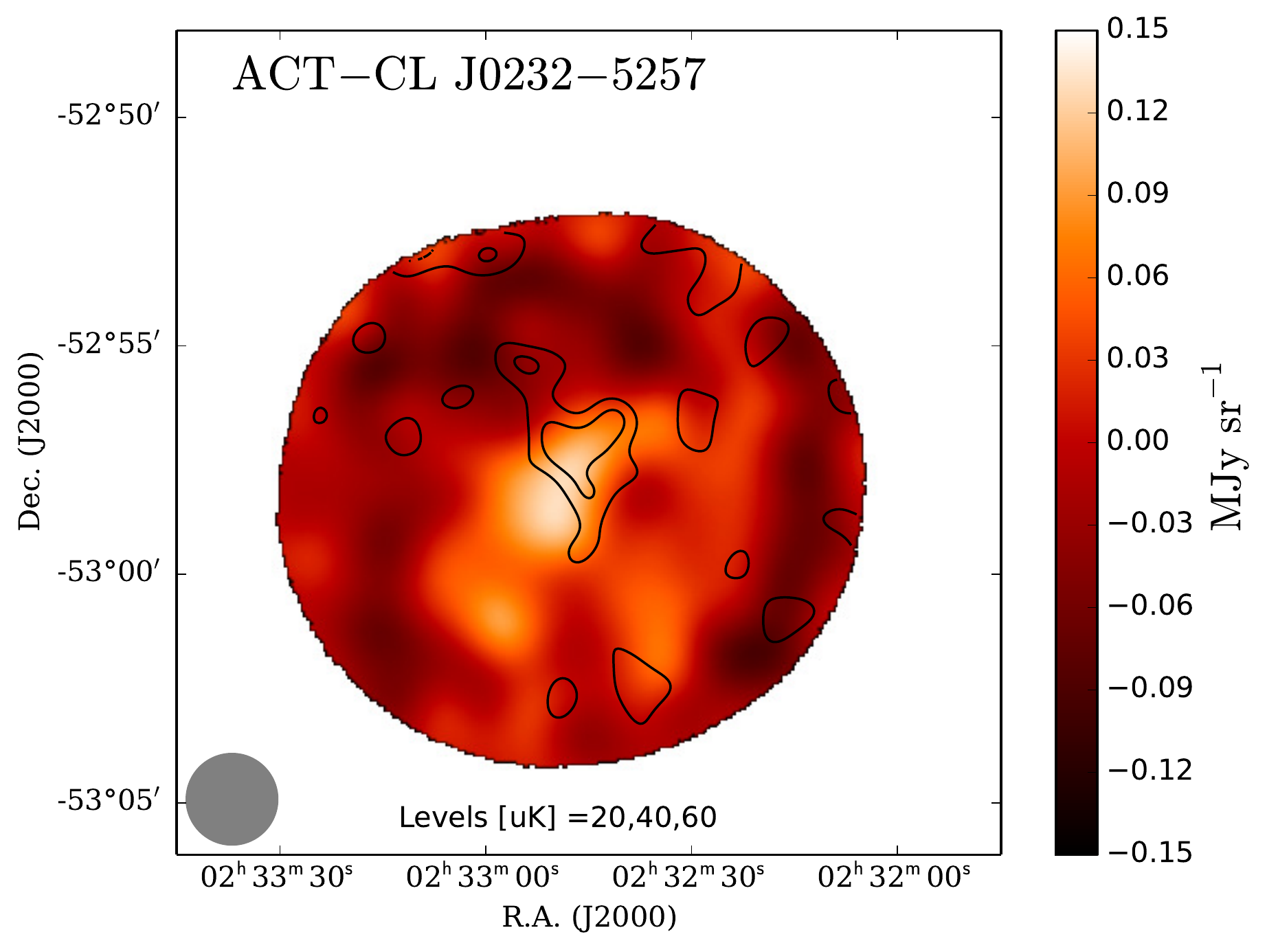} &
  \includegraphics[scale=0.45]{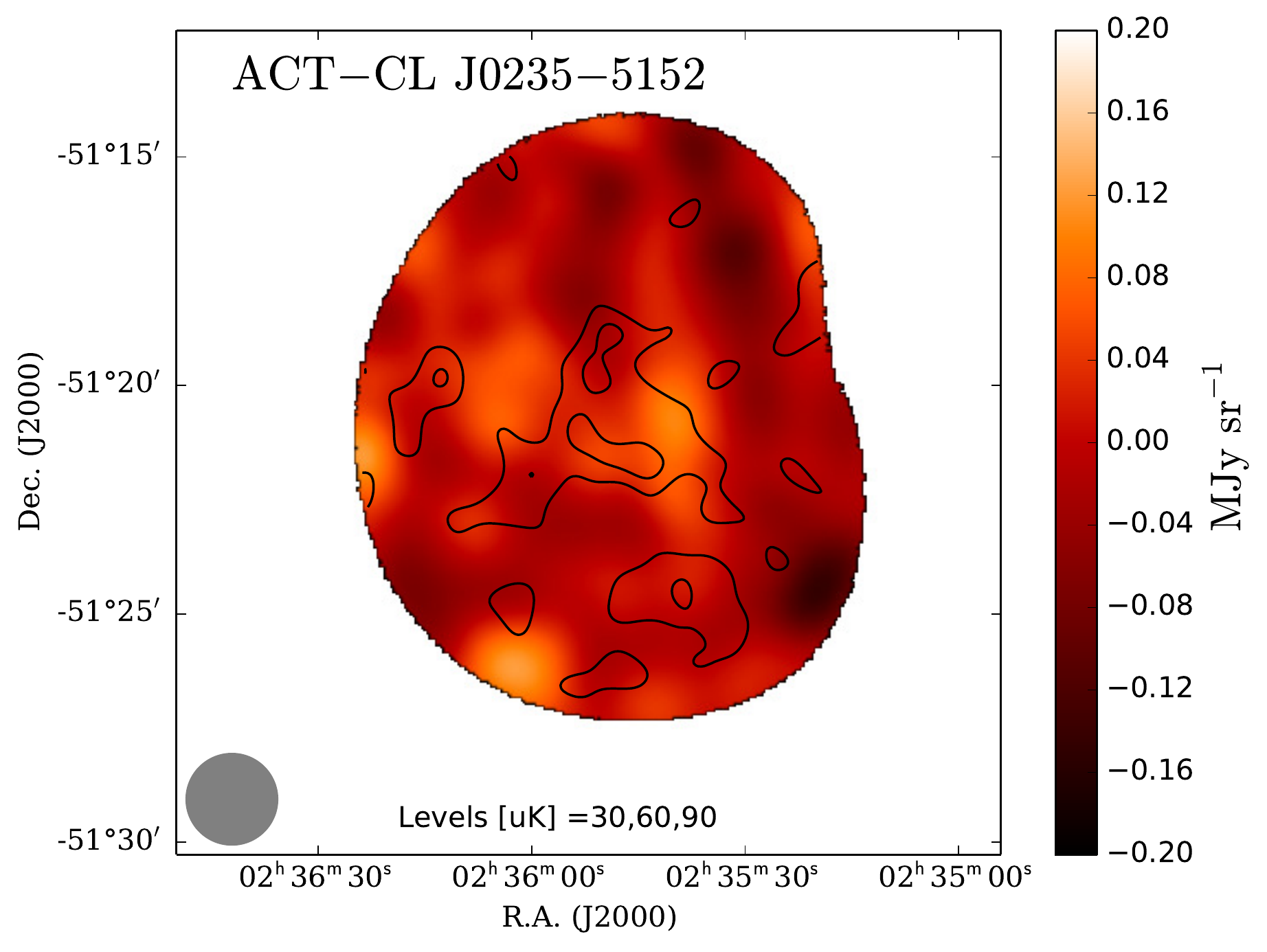} \\
  \includegraphics[scale=0.45]{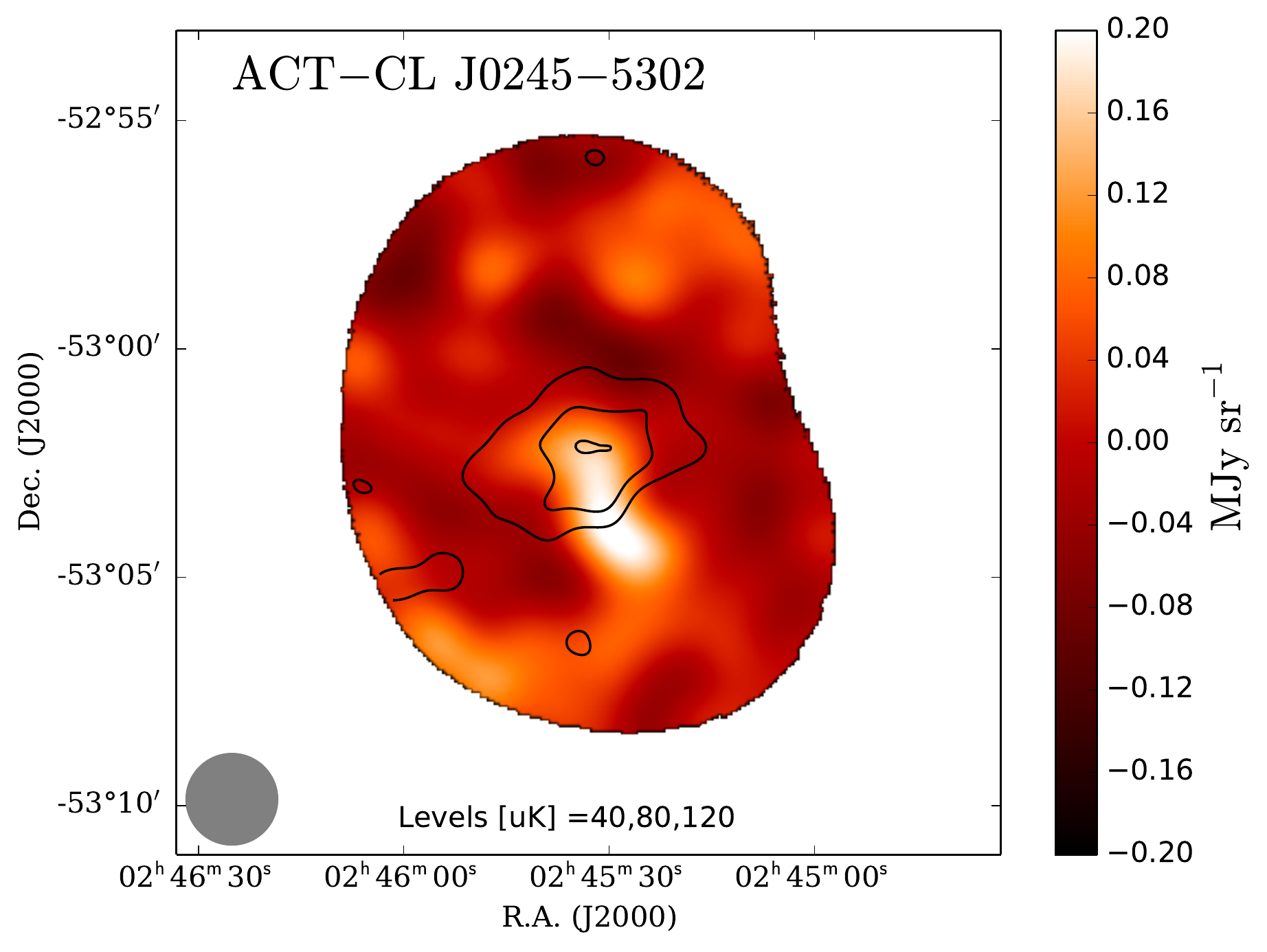} &
  \includegraphics[scale=0.45]{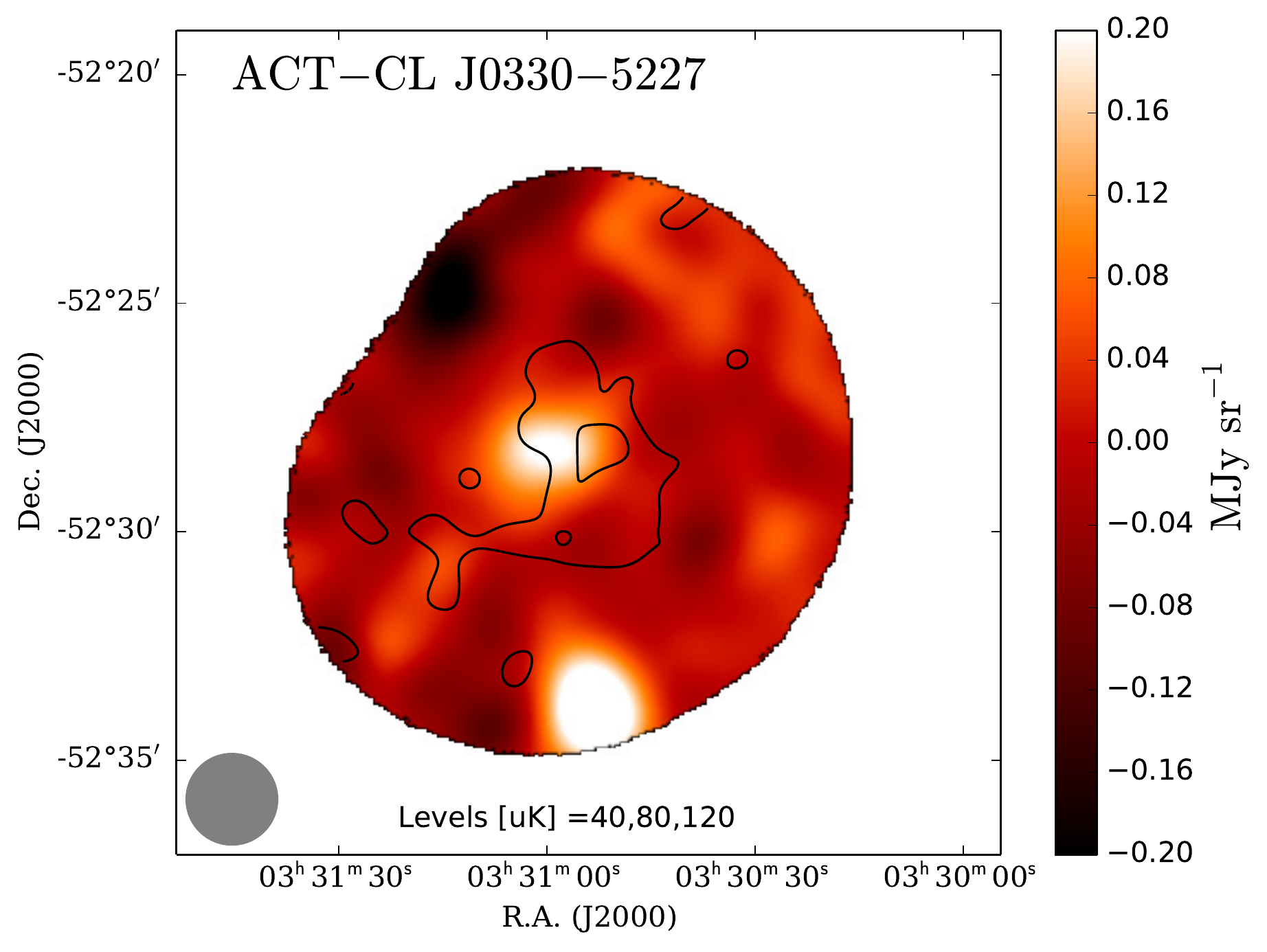}
\end{tabular}
\caption[LABOCA $345\,\rm GHz$ maps]
{LABOCA $345\,\rm GHz$ maps after iterative
filtering, point-source subtraction, and smoothing.  
The color scale images represent $345\,\rm GHz$ surface 
brightness.
The black contours represent the 
LABOCA-filtered ACT\,$148\,\rm GHz$ decrement intensities 
with levels indicated in the panels.
The grey circle at the lower left represents the effective
beam size of the LABOCA maps after smoothing to match
the ACT resolution ($1.4^{\prime}$).}
\label{f-345_images}
\end{figure*}
\addtocounter{figure}{-1}
\begin{figure*}
\centering
\begin{tabular}{cc}
  \includegraphics[scale=0.45]{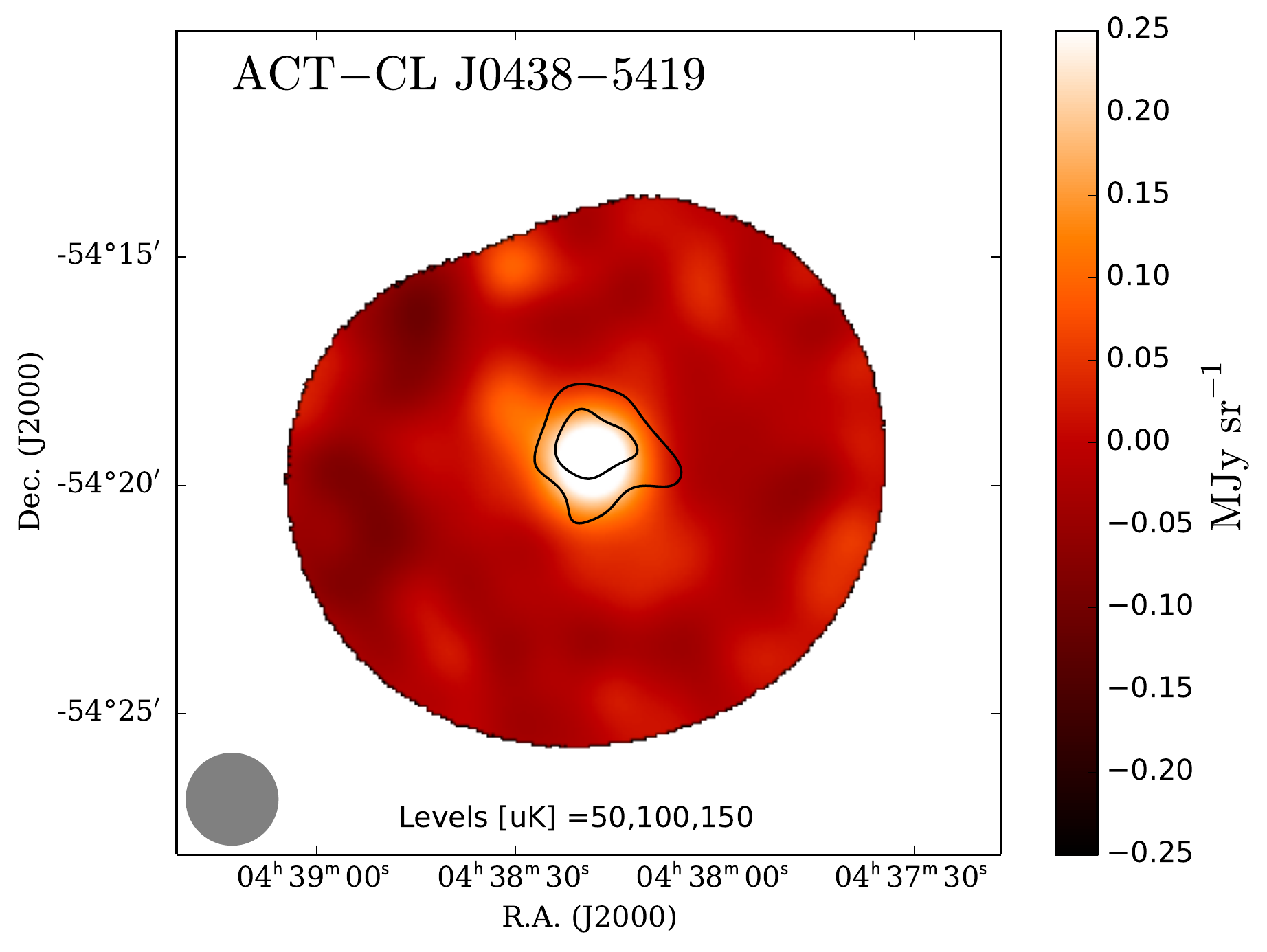} &
  \includegraphics[scale=0.45]{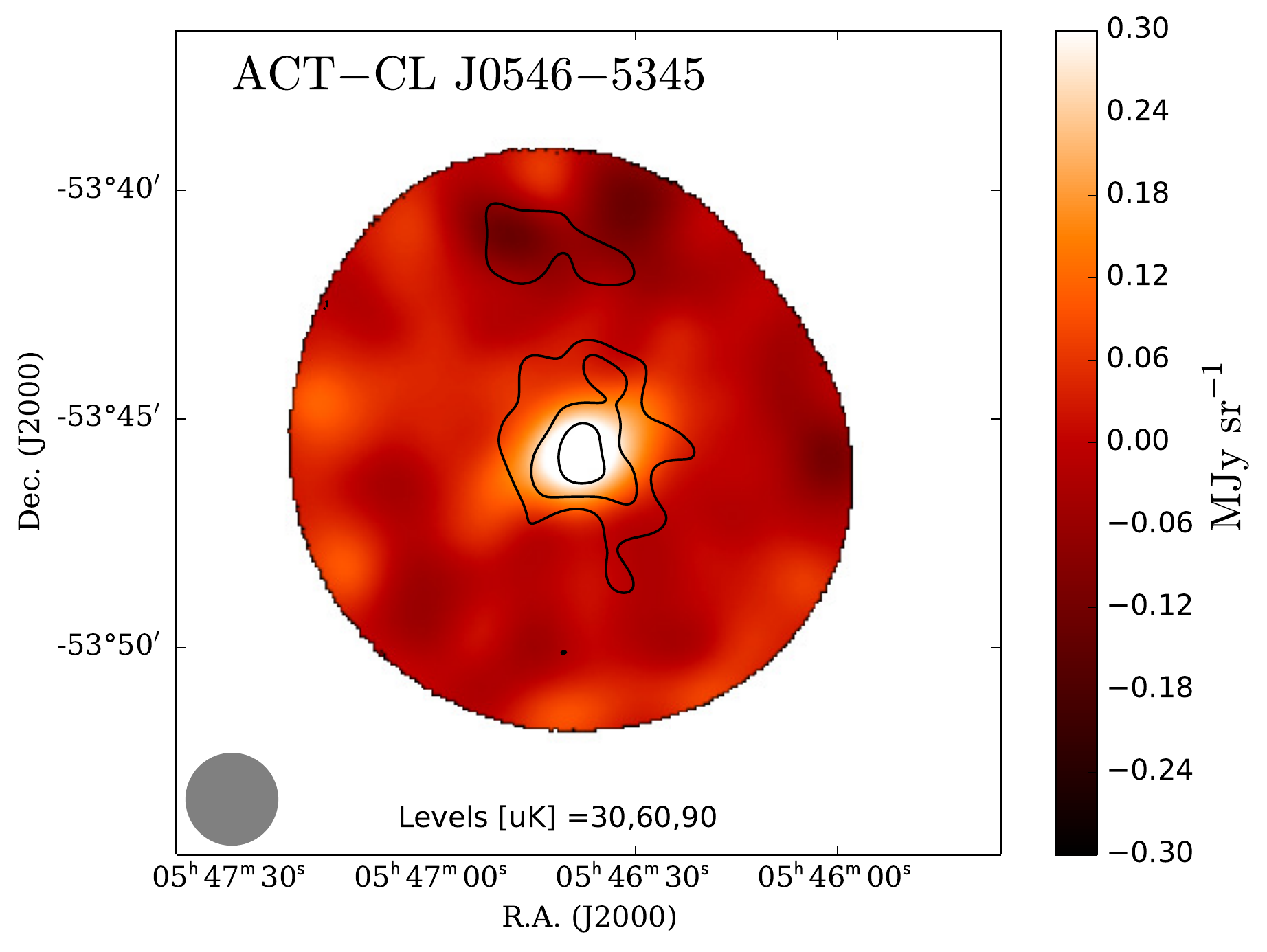} \\
  \includegraphics[scale=0.45]{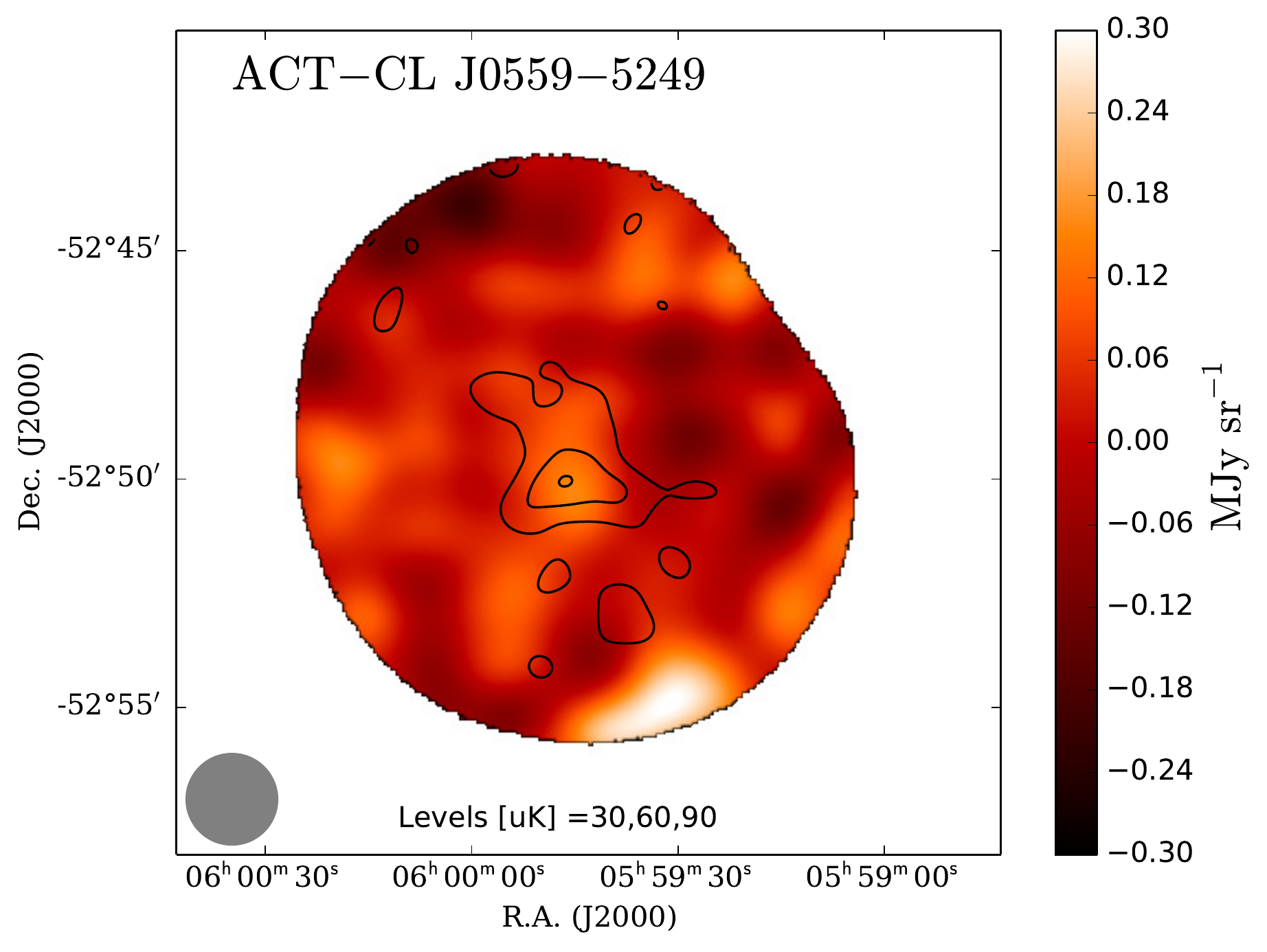} &
  \includegraphics[scale=0.45]{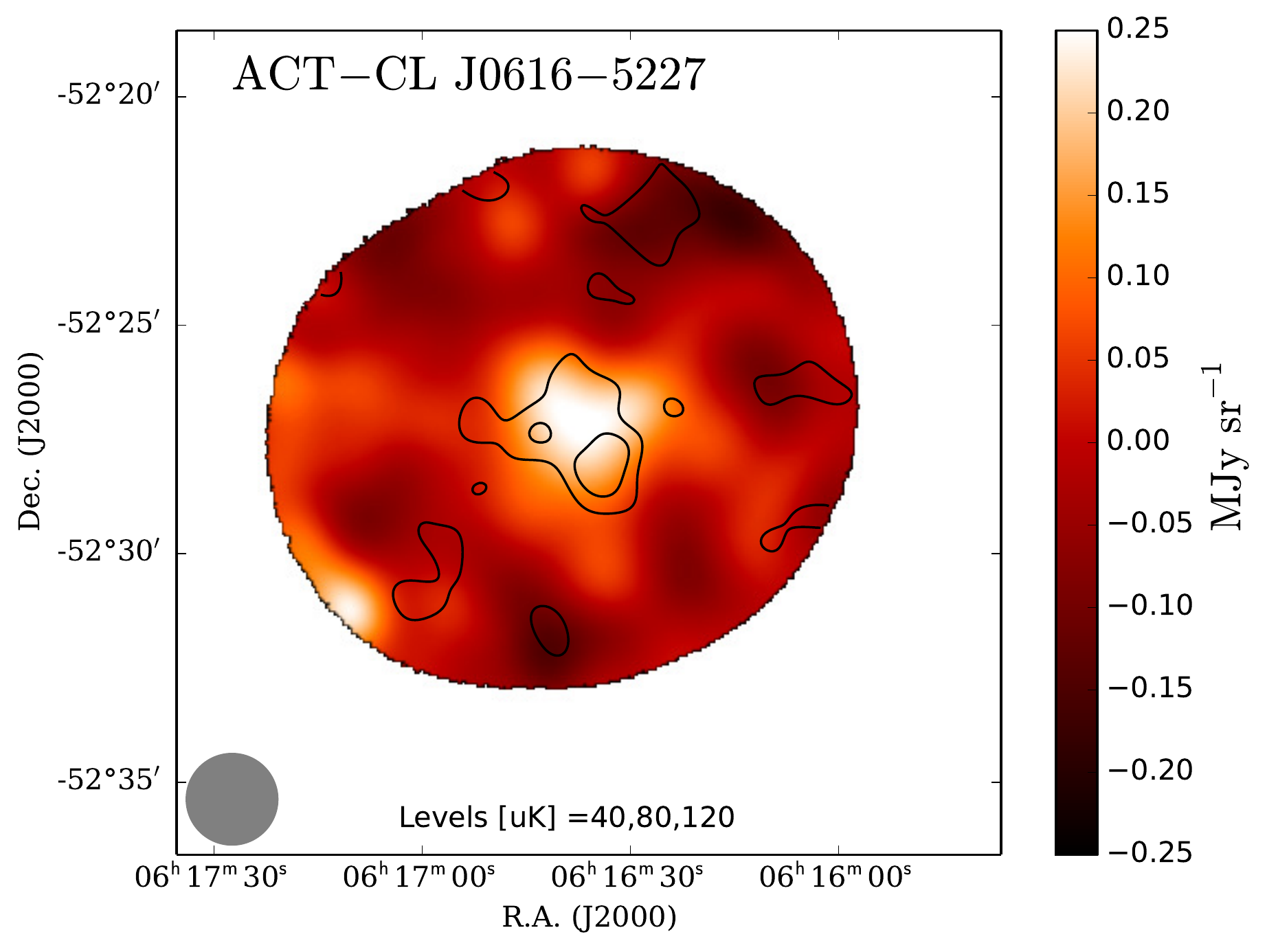} \\
  \includegraphics[scale=0.45]{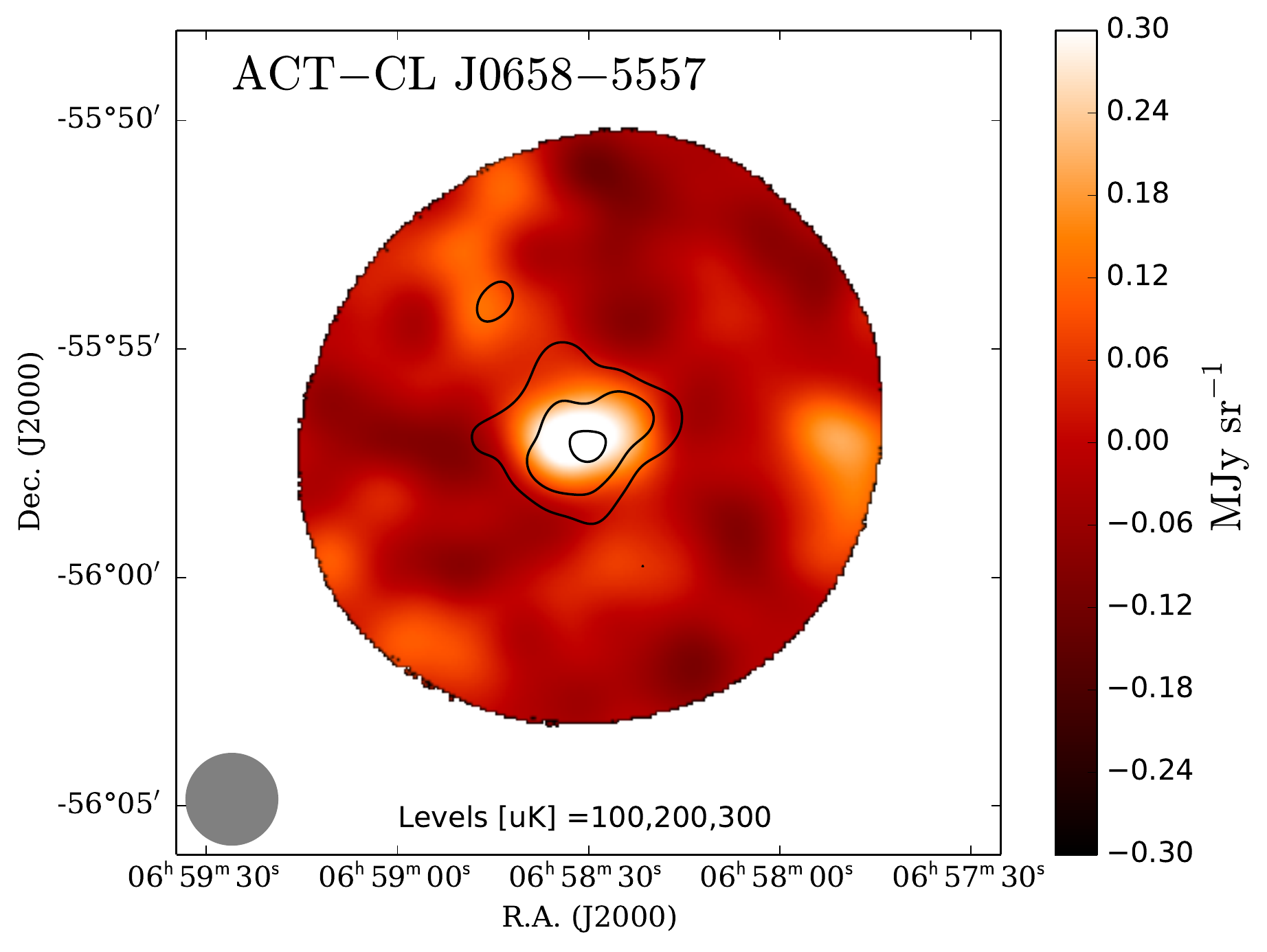}
\end{tabular}
\caption[]{Continued.}
\end{figure*}
Observations were taken using the standard 
raster spiral mode, in which the telescope traces out
one spiral every 35 seconds at each 
of four raster points defining a square with 
$27^{\prime\prime}$ sides. 
In polar coordinates $(r,\phi)$, the spiral track has an initial 
radius $r_{0}=18^{\prime\prime}$, a constant radial speed
$\dot{r}=2.2^{\prime\prime}\,\rm s^{-1}$, and a constant 
angular rate $|\dot{\phi}|=\frac{\pi}{2} \,\rm rad\,s^{-1}$. 
The $345\,\rm GHz$ zenith  
opacity is interpolated between skydip measurements that 
punctuate observing sessions, and is used to correct for
line-of-sight atmospheric absorption. 
Flux calibration is determined by observing a planet
or secondary 
calibrator\footnote{http://www.apex-telescope.org/bolometer/laboca/calibration/\label{laboca-cal}} 
before each observing session, and the telescope pointing is monitored 
throughout the observations with periodic scans of 
bright quasi-stellar objects (QSOs).  
Scans that either had abnormally high RMS noise or 
were taken during rapidly changing atmospheric conditions 
were removed from the analysis.  The total 
on-target integration time for the sample is
140\,hr, as shown in Table \ref{t-clusters}.

The LABOCA main beam has a Gaussian profile with
full width at half maximum (FWHM) of $19.2^{\prime\prime}$.  The full 
beam$^{\ref{laboca-cal}}$ includes broader, low-level wings and has total
$\Omega = 12.17\pm 0.43\,\rm nsr$.
We use the full beam   
when computing integrated flux densities of extended
sources, and we use the Gaussian main lobe when fitting
PSF profiles to point sources.

\subsubsection{LABOCA data reduction}
\label{sss-laboca-reduction}
We reduced the LABOCA data using the Python-based  
Bolometer Array Analysis
Software 
(BoA\footnote{http//:www.apex-telescope.org/bolometer/laboca/boa/}) 
package.  The data time-stream $T(c,t)$, a function of
channel $c$ and time $t$, is first flux 
calibrated and corrected for atmospheric absorption. Next, 
median filtering
at each time step is applied to all channels, then to
each of twelve subgroups of channels that share 
readout cabling, then to each 
of four subgroups that 
share amplifier boxes.
Despiking is performed between stages of median filtering, 
before a linear baseline is subtracted from each channel; 
channels with RMS noise greater than $4\times$ the median value 
are flagged.  Low-frequency ``$1/f$'' noise is removed using 
BoA's noise whitening algorithm \texttt{flattenFreq}, 
which sets the magnitudes of Fourier modes with 
frequencies less than some cutoff frequency $f<f_c$ to the 
average magnitude of those in the range $f_c < f < 1.2\times f_c$.
Because the celestial scanning velocity increases as a function 
of time during each scan as 
$v(t)\simeq(r_0+\dot{r}t) |\dot{\phi}|$,
\texttt{flattenFreq} filters out emission on angular scales larger than 
the cutoff scale $s_c(t)$ given by\footnote{In Equation 
\ref{e-lmax}, the term $\dot{r}^2$ 
in the full equation for the scanning velocity, 
$v^2=\dot{r}^2 + (r_0+\dot{r}t)^2|\dot{\phi}|^2$, 
is ignored because $\dot{r}^2 \ll (r_0+\dot{r}t)^2|\dot{\phi}|^2$
during the entire scan.}
\begin{align}
\label{e-lmax}
s_c(t) & =v(t)/f_c =\left( 28 + 3.5\times t
\right) \, f_c^{-1} \; [\rm arcsec],
\end{align}
with $t$ in seconds and $f_c$ in 
Hz.
The data are then
gridded onto an equatorial $0.3^{\circ}\times 0.3^{\circ}$ image with 
$3.6^{\prime\prime}$ pixels (oversampling the beam by a factor
of five in each direction).  The resulting RMS sensitivities 
within $4^{\prime}$ of the centers of the beam-smoothed maps
are presented in Table \ref{t-clusters}.
\begin{deluxetable*}{ccccccccc}
\tablecaption{Cluster sample}
\tablewidth{0pt}
\tablehead{
\colhead{} & 
\colhead{R.A.\tablenotemark{a}} & 
\colhead{Dec.\tablenotemark{a}} &
\colhead{} &
\colhead{$\theta_{500c}$} &
\colhead{$M_{500c}$} &
\colhead{} & 
\colhead{$t_{\rm obs}$\tnm{b}}&
\colhead{RMS\tnm{c}} 
\\
\colhead{Name} & 
\colhead{(h:m:s)} & 
\colhead{($\circ:\prime:\prime\prime$)} &
\colhead{$z$} &
\colhead{($\prime$)}&
\colhead{($10^{14}M_{\odot}$)}&
\colhead{Project ID (P.I.)} & 
\colhead{(hr)}&
\colhead{($\rm mJy\,bm^{-1})$}
}
\startdata
ACT-CL\,J0102$-$4915 &  01:02:53 & -49:15:19 & 0.870\tnm{d} & 2.50\tnm{e}  & $10.2 \pm 2.4$\tnm{e} & M-087.F-0037-2011   (A. Baker)      &   11.3 & 2.4 \\
ACT-CL\,J0215$-$5212 &  02:15:18 & -52:12:30 & 0.480\tnm{e} & 3.16\tnm{e}  & $ 6.0 \pm 1.7 $\tnm{e} & C-088.F-1772A-2011  (L. Infante)    &   17.6 & 2.0 \\
ACT-CL\,J0232$-$5257 &  02:32:45 & -52:57:08 & 0.556\tnm{e} & 2.42\tnm{e}  & $ 3.7 \pm 1.4 $\tnm{e} & O-086.F-9302A-2010  (A. Baker)      &   17.0 & 1.7  \\
ACT-CL\,J0235$-$5121 &  02:35:52 & -51:21:16 & 0.278\tnm{e} & 5.18\tnm{e}  & $ 7.4 \pm 2.1 $\tnm{e} & O-087.F-9300A-2011  (A. Baker)      &   12.2 & 2.1 \\
ACT-CL\,J0245$-$5302 &  02:45:33 & -53:02:04 & 0.300\tnm{f} & 3.08\tnm{g}  & $ \simeq 3.1 $\tnm{g} & M-088.F-0003-2011   (A. Wei\ss)     &   11.6 & 2.0 \\
ACT-CL\,J0330$-$5227 &  03:30:54 & -52:28:04 & 0.442\tnm{e} & 4.08\tnm{e}  & $ 10.7 \pm 2.5 $\tnm{e} & E-086.A-0972A-2010  (A. Baker)      &   8.1  & 1.9 \\
ACT-CL\,J0438$-$5419 &  04:38:19 & -54:19:05 & 0.421\tnm{e} & 4.53\tnm{e}  & $ 13.2 \pm 3.1 $\tnm{e} & E-086.A-0972A-2010  (A. Baker)      &   18.3 & 1.6 \\
ACT-CL\,J0546$-$5345 &  05:46:37 & -53:45:32 & 1.066\tnm{e} & 1.75\tnm{e}  & $ 5.1 \pm 2.6 $\tnm{e} & C-086.F-0668A-2011  (L. Infante)    &  16.3 & 1.6 \\
ACT-CL\,J0559$-$5249 &  05:59:43 & -52:49:13 & 0.609\tnm{e} & 3.08\tnm{e}  & $9.3 \pm 2.7 $\tnm{e} & C-087.F-0012A-2011  (L. Infante)    &   13.3 & 2.8 \\
ACT-CL\,J0616$-$5227 &  06:16:36 & -52:28:04 & 0.684\tnm{e} & 2.60\tnm{e}  & $7.0 \pm 3.1 $\tnm{e} & O-088.F-9300A-2011  (A. Baker)      &   14.8 & 2.1 \\\hline
\hline
\multirow{2}{*}{ACT-CL\,J0658$-$5557} &  \multirow{2}{*}{06:58:30} & \multirow{2}{*}{-55:57:04} & \multirow{2}{*}{0.296\tnm{h}} & \multirow{2}{*}{3.44\tnm{i}}  & \multirow{2}{*}{$11.0 \pm 6.8$\tnm{i}} & E-380-A-3036A-2007 (M. Birkinshaw) &   \multirow{2}{*}{16.5}  & \multirow{2}{*}{2.2} \\
                     &           &           &   &           &              & O-079.F-9304A-2007 (D. Johansson) &    
\enddata
\tablenotetext{a}{SZE decrement centroid from \citet{marr11}}
\tablenotetext{b}{Total un-flagged, on-target integration time}
\tablenotetext{c}{RMS in the beam-smoothed maps, which have an effective
resolution of $28^{\prime\prime}$.}
\tablenotetext{d}{\citet{mena12}}
\tablenotetext{e}{\citet{sifo13}}
\tablenotetext{f}{\citet{edge94}}
\tablenotetext{g}{$r_{500c}$ for ACT-CL\,J0245$-$5302
was estimated using its velocity dispersion 
$\sigma\simeq 900 \,\rm km\,s^{-1}$ \citep{edge94}, giving
$M_{200c}\sim 5\times 10^{14}\,M_{\odot}$ 
and $r_{200c}=1260\,\rm kpc$.}
\tablenotetext{h}{\citet{tuck98}}
\tablenotetext{i}{$r_{500c}$ from \citet{zhan06}}
\label{t-clusters}
\end{deluxetable*}

The above reduction steps, represented by $\mathcal R$, 
operate on time-stream data and return a gridded map, i.e.,
$I(\alpha,\delta)=\mathcal R[T(c,t),l]$, where the parameter 
$l$ represents the largest source size to which the entire scan 
remains responsive.  
The corresponding $f_c$ is found from Equation
\ref{e-lmax} by setting $s_c$ at $t=0$ (where it achieves
its minimum value) equal to $l$ convolved with the 
LABOCA beam FWHM $\theta_{\rm beam}$:
\begin{equation}
  f_c(l) = \frac{28}{\sqrt{l^2+\theta_{\rm beam}^2}} [\rm Hz],
\end{equation}
for $l$ and $\theta_{\rm beam}$ in arcsec.  For 
$l=0^{\prime\prime}$ and $l=120^{\prime\prime}$, 
$f_c(l) = 1.47$ and $0.23\rm \,Hz$, respectively.

\subsubsection{Iterative multi-scale algorithm}
\label{s-iterative}
\begin{figure*}
\centering
\includegraphics[scale=0.8]{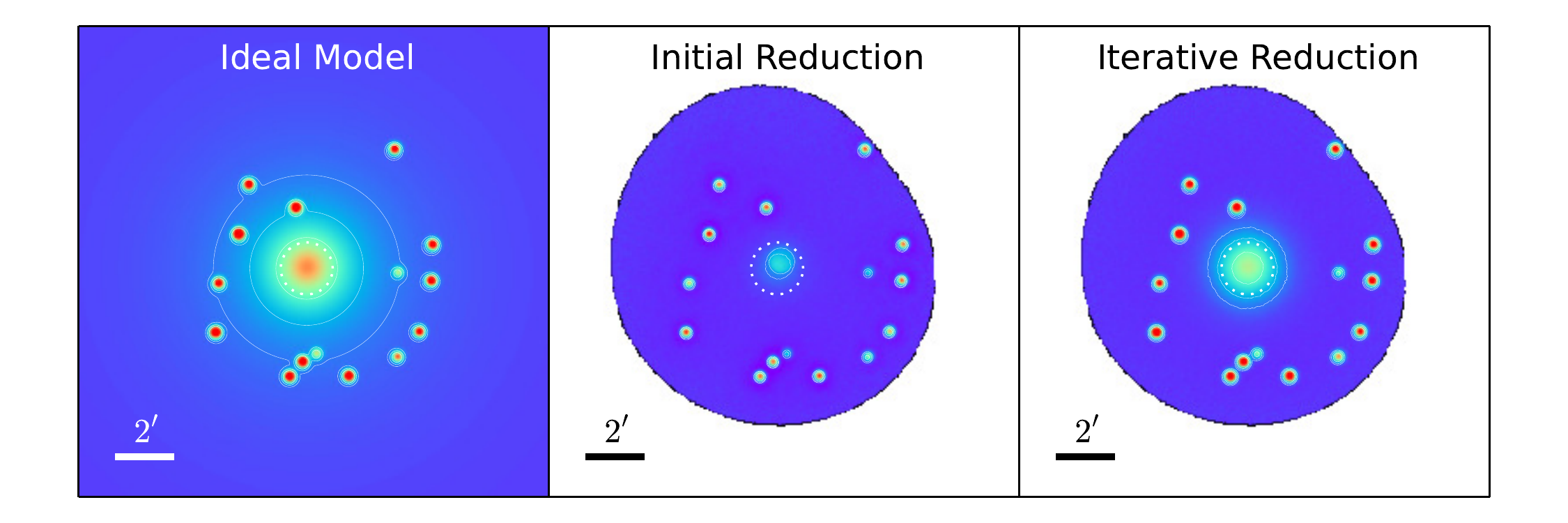}
\caption[LABOCA iterative pipeline performance]
{Example of iterative reduction performance on an
ideal cluster field.  The cluster SZE signal is modelled
using an iso-thermal beta profile with $\beta=0.86$ and core
radius $r_{\rm c}=1^{\prime}$.  SMGs are modeled as 15 point 
sources injected at random locations. The reduction adopts
the observational parameters from 
ACT-CL\,J0546$-$5345. {\em Left:} Input $345\,\rm GHz$ 
noise-free model.  {\em Center:}  Output image after a 
single reduction (no iteration, Section 
\ref{sss-laboca-reduction}).  {\em Right:} 
Output image after running the iterative pipeline 
(described in Section \ref{s-iterative}).
The white dotted dashed circle is centered on the modelled 
cluster with radius equal to the modeled core radius.  
White solid contours represent $345\,\rm GHz$ intensity levels of 
1, 2, and $4\,\rm mJy\,beam^{-1}$.  The fraction of recovered
SZE signal within a $1.5^{\prime}$ aperture is initially
\RawSZEEfficiency{}, increasing to \IterativeSZEEfficiency{} 
after running the iterative multi-scale pipeline.}
\label{f-trans_image}
\end{figure*}

The filtering steps described above, 
primarily median filtering on cabling subgroups
and noise whitening (Section 
\ref{sss-laboca-reduction}), can remove the desired astrophysical 
emission from the data along with atmospheric noise.  
To recover this lost signal, we have developed an 
iterative, multi-scale pipeline. Our approach is inspired
by techniques used by 
\citet{enoc06} and \citet{nord09}.
One important 
difference is that we maximize our sensitivity to 
low-level extended emission by using a series of 
matched filters to search for signal at multiple 
cluster-sized spatial 
scales, similar in nature to 
adaptive filtering algorithms \cite[e.g., ][]{scov07}. 

In the technique described below, a single ``reduction''
refers to one application of the standard pipeline
described in Section \ref{sss-laboca-reduction}.  Our 
iterative pipeline is built ``on top'' of the standard
pipeline by chaining together standard reductions using
different values of $f_c$.
The first step of the pipeline is reducing the data with
aggressive filtering by setting the angular scale of 
fully preserved emission to $l=0^{\prime\prime}$. 
This first image, $I^{l=0^{\prime\prime}}_1$, is minimally affected 
by $1/f$ noise, although it has complete 
responsiveness only to point sources. 
High-significance structures in this first image are 
located by producing a S/N 
image $S(\alpha,\delta)$ using a spatial matched 
filter \citep[e.g., ][]{serj03}
\begin{equation}
S  = \frac{I \otimes k}{\sqrt{W \otimes k^2}},
\end{equation}
where $k$ is a 2D Gaussian kernel with 
${\rm FWHM} = \sqrt{l^2+\theta_{\rm beam}^2}$, and
$W(\alpha,\delta)$ is a weight map defined by the 
inverse variance of time stream data within each pixel.
Pixels where $S(\alpha,\delta)<S_{\rm thresh}$ are
set to zero in $I^{l=0^{\prime\prime}}_{1}$.
The significance level $S_{\rm thresh}$ 
is chosen so that we would expect $\leq 0.3$ spurious sources in the
S/N map due to noise fluctuations (assuming a Gaussian noise 
distribution), and therefore depends on $l$ and the 
image size.
This clipped image is then smoothed to the angular
scale of interest by convolving with a normalized 
2D Gaussian with ${\rm FWHM}=l$ to produce a model 
image $M^{l=0^{\prime\prime}}_1$. For the case of 
$l=0^{\prime\prime}$, a small amount of smoothing, 
$\rm FWHM=7^{\prime\prime}$, is still applied to help 
reduce sharp discontinuities in the model image.  
If the model image $M_1^{l=0^{\prime\prime}}$ 
is not empty, iteration 
proceeds and $M_1^{l=0^{\prime\prime}}$ 
is transformed into time stream data 
${\mathcal T}[M_1^{l=0^{\prime\prime}}]$ and subtracted from 
the original time stream. This residual time stream 
$T-{\mathcal T}[M_1^{l=0^{\prime\prime}}]$ is then 
reduced using the procedure from Section \ref{sss-laboca-reduction},
and the resulting residual image is added to $M_1^{l=0^{\prime\prime}}$ 
to produce the next image, 
$I_2={\mathcal R}[T-{\mathcal T}[M_1]]+M_1$.
This process is carried out until the signal in the
map converges, ($\Sigma (I_{n}-I_{n-1})\simeq 0$), 
and we are left with a final image $I_{N}$ and 
model $M_{N}$.  Ten iterations are sufficient for
the model image to converge. 

Now we begin to search for larger spatial scale emission.
$M_N^{l=0^{\prime\prime}}$ (the final, converged model image with $l=0^{\prime\prime}$
filtering) is subtracted 
from the original time streams, and the residual time stream
is reduced with a relaxed filtering, initially, 
$l=30^{\prime\prime}$.
If high-significance $30^{\prime\prime}$-scale
emission is located using a matched filter, 
iteration begins again.
This process is carried out for
$l=0^{\prime\prime},30^{\prime\prime},60^{\prime\prime},$ and
$120^{\prime\prime}$; the final image is the sum of all 
converged models, plus any residual low-level signal and the
remaining noise:
\begin{equation}
I_{\rm final} = M_f + {\mathcal R}(T-{\mathcal T}[M_f]),
\end{equation}
where $M_f$ is the sum of all scales' converged model 
images  $M_f=\Sigma_{l}{M_N^l}$.
The array loses sensitivity quickly for angular scales
that are larger than the typical size of contiguous 
subsets of channels that share cabling 
(i.e., $\gtrsim 6.7^{\prime}$); 
therefore, the pipeline is limited to 
$l\leq 120^{\prime\prime}$,
which allows for recovery of emission on scales up to
$\left<s_c(t,l=120^{\prime\prime})\right>\leq 6.5^{\prime}$.

\begin{deluxetable*}{cccclclc}
\tablecaption{SZE properties}
\tablewidth{0pt}
\tablehead{
\colhead{} & 
\colhead{$T_{e}$} &
\colhead{$\theta^{\prime}$\tnm{a}} &
\colhead{} &
\colhead{$v_p$\tnm{b}} &
\colhead{$S^{\rm SZE}_{\rm 345}(<\theta^{\prime})$} &
\colhead{$Y^{\prime}_{\rm SZ}$\tnm{b,c}} &
\colhead{} 
\\
\colhead{Name} &
\colhead{$(\rm keV)$}&
\colhead{$(\rm arcmin)$}&
\colhead{SPIRE?\tnm{b}} &
\colhead{$(\rm km\,s^{-1})$} &
\colhead{(mJy)} &
\colhead{($10^{-10}\,\rm sr$)} &
\colhead{S/N\tnm{d}} 
}
\startdata
 $\rm S/N \ge 5_{}$ \\
\tableline
ACT-CL\,J0102$-$4915 & $14.5\pm 1.0$\tnm{e}          &  2.2 &  \checkmark & $-1100_{- 2200}^{+ 1300}$ ($-2800_{- 3100}^{+ 1700}$)  &  $ 276 \pm  28$  &  $ 1.8_{- 0.3}^{+ 0.3}$ ($ 1.4_{- 0.3}^{+ 0.2}$)  &  9.7 \\
ACT-CL\,J0215$-$5212 & $5.9\pm1.3$\tnm{f}            &  1.9 &           & $-1100_{-  600}^{+  400}$  &  $ 171 \pm  21$  &  $ 0.8_{- 0.1}^{+ 0.1}$  &  8.2 \\
ACT-CL\,J0438$-$5419 &  $11.9\pm 1.2$\tnm{f}         &  1.9 &\checkmark & $  900_{- 1000}^{+ 1000}$ ($  600_{- 1200}^{+ 1000}$)  &  $ 127 \pm  15$  &  $ 0.9_{- 0.1}^{+ 0.1}$ ($ 0.8_{- 0.1}^{+ 0.1}$)  &  8.7 \\
ACT-CL\,J0546$-$5345 & $8.54^{+1.38}_{-1.05}$\tnm{g}    &  1.9 &\checkmark & $ -300_{-  700}^{+  700}$ ($ -300_{-  700}^{+  700}$)  &  $ 162 \pm  20$  &  $ 1.0_{- 0.1}^{+ 0.1}$ ($ 0.9_{- 0.1}^{+ 0.1}$) &  8.2 \\
ACT-CL\,J0616$-$5227 & $6.6\pm 0.8$\tnm{f}           &  1.9 &           & $ -300_{-  600}^{+  600}$  &  $ 160 \pm  21$  &  $ 0.9_{- 0.1}^{+ 0.1}$  &  7.7 \\
ACT-CL\,J0658$-$5557 & $10.8\pm 0.9$\tnm{h}          &  1.6 &            &$ 2500_{- 1000}^{+ 1000}$  &  $ 121 \pm  22$  &  $ 1.1_{- 0.2}^{+ 0.1}$  &  5.4 \\
\tableline
 $\rm S/N<5_{}$ \\
\tableline
ACT-CL\,J0232$-$5257 & $9.1\pm 2.1$\tnm{f}           &  1.8 &            &$-1200_{- 1900}^{+ 1200}$  &  $  67 \pm  16$  &  $ 0.4_{- 0.1}^{+ 0.1}$  &  4.3 \\
ACT-CL\,J0235$-$5121 & 5.0\tnm{c}                    &  1.7 & \checkmark & $-1500_{- 2800}^{+ 1800}$ ($-1500_{- 2800}^{+ 1800}$)  &  $  36 \pm  16$  &  $ 0.2_{- 0.1}^{+ 0.1}$ ($ 0.2_{- 0.0}^{+ 0.1}$)  &  2.2 \\
ACT-CL\,J0245$-$5302 & 5.0\tnm{c}                    &  1.9 & \checkmark & $  500_{-  600}^{+ 1200}$ ($  500_{-  600}^{+ 1200}$)  &  $  53 \pm  18$  &  $ 0.4_{- 0.1}^{+ 0.1}$ ($ 0.4_{- 0.1}^{+ 0.1}$)  &  2.9 \\
ACT-CL\,J0330$-$5227 & $4.32^{+0.21}_{-0.19}$\tnm{g}    &  1.9 & \checkmark & $  100_{- 1000}^{+ 1000}$  ($  100_{-  800}^{+ 1000}$) &  $  70 \pm  34$  &  $ 0.5_{- 0.2}^{+ 0.2}$ ($ 0.5_{- 0.2}^{+ 0.1}$)  &  2.1 \\
ACT-CL\,J0559$-$5249 & $8.09\pm 0.75$\tnm{g}         &  1.5 &            &$ 3100_{- 1000}^{+ 1900}$  &  $  41 \pm  23$  &  $ 0.4_{- 0.1}^{+ 0.1}$  &  1.8 
\enddata
\tablenotetext{a}{Aperture radius used in computing $Y^{\prime}_{\rm SZ}$}
\tablenotetext{b}{{\em Herschel} SPIRE data obtained through proposal OT2\_abaker01\_2.  Results that include SPIRE data
are shown in parentheses.}
\tablenotetext{c}{For clusters with unknown $T_e$, $Y_{\rm SZ}^{\prime}$ is
computed assuming $T_e=5.0\,\rm keV$.}
\tablenotetext{d}{S/N based on integrated 
                   $S_{345}$ inside $Y^{\prime}_{\rm SZ}$ apertures
                   using the point-source subtracted, iteratively reduced maps.}
\tablenotetext{e}{\citet{mena12}}
\tablenotetext{f}{Hughes et al., in prep.}
\tablenotetext{g}{\citet{mena10}}
\tablenotetext{h}{Mass-weighted temperature from \citet{halv09}}
\label{t-sze-signal}
\end{deluxetable*}

We detect bright SMGs in all clusters and 
strong SZE increment signals (S/N$>3.5$) in 
six clusters (see Table \ref{t-sze-signal}).
The final, converged, iteratively reduced
$345\,\rm GHz$ images of all eleven clusters
after point-source subtraction and smoothing
are shown in Figure \ref{f-345_images}.

\begin{figure}
\centering
\includegraphics[scale=0.45]{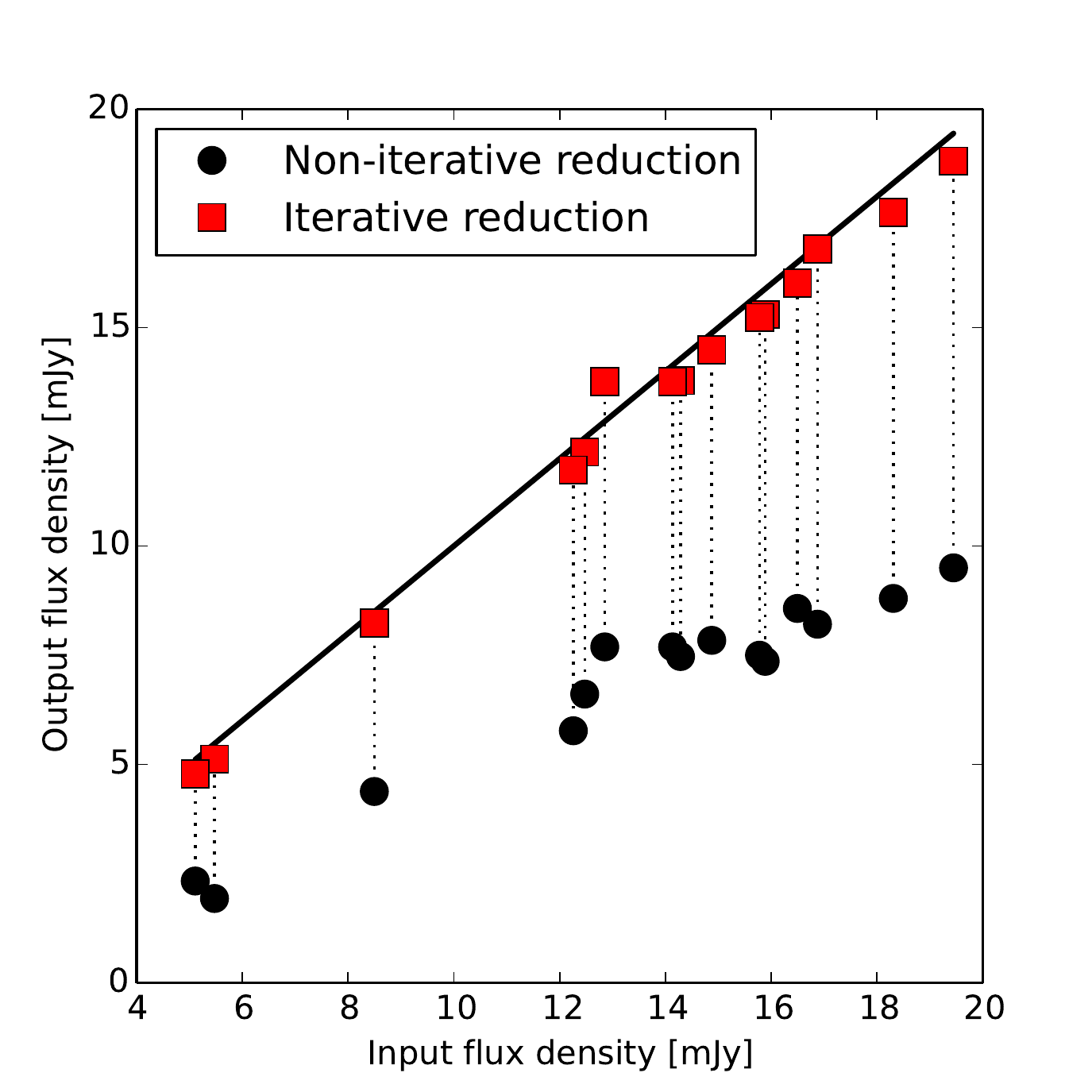}
\caption[Iterative pipeline transfer function efficiency]
{Point source transfer function efficiency of the iterative 
LABOCA reduction pipeline.  The circles and squares represent 
the input and output flux densities of 15 ideal LABOCA point sources
with flux densities between 5--$19\,\rm mJy$ (see Figure 
\ref{f-trans_image}) for a single run of the reduction pipeline 
(black circles), and for a full iterative reduction (red squares).
The black line shows  the relation $\rm output= input$. 
The average flux recovery fractions for single and iterative reductions
are \RawEfficiency{} and \IterativeEfficiency{}, respectively.} 
\label{f-trans_phot}
\end{figure}
As an instructive check of our approach to data processing, we
can compare our map of ACT\,J0658-5557 to the 345\,GHz map made by
\citet{joha10} from the {\it same} LABOCA data using different
reduction software (CRUSH\footnote{http://www.submm.caltech.edu/~sharc/crush/}), and to
the $1^\prime$ resolution 150\,GHz map of the cluster made with
the APEX-SZ instrument.  \citet{joha10} deliberately
filter the LABOCA data to suppress extended structure, including
SZE increment signal.  However, the flux densities 
they measure for the three most significantly detected SMGs in their
Gaussian-matched-filtered map ($48.6 \pm 1.3$, $15.1 \pm 1.0$,
and $6.9 \pm 0.9\,{\rm mJy}$) agree well with those we have
extracted from our own version of the map ($48 \pm 2$, $13 \pm 2$,
and $7 \pm 2\,{\rm mJy}$: Aguirre et al., in preparation).
\citet{halv09} fit an isothermal elliptical $\beta$
model to their SZE decrement observations and determine a projected
core radius of $142 \pm 18\,{\rm arcsec}$ that is well-matched to
the size of the (rather asymmetric) structure we see in
our 345\,GHz map.  Broadly speaking, our mapping algorithm 
produces results that are consistent with previous work 
that focuses on either point sources or extended emission.

\subsubsection{Systematic uncertainties of iterative pipeline}
\label{ss-transfer}
In Figure \ref{f-trans_image}, we show the results when
a noise-free\footnote{Our ``noise-free'' data
actually have a non-zero RMS noise of $1\%$ that 
of the real data, due to constraints of the
BoA software package.  Realistic correlations in the 
simulated noise are produced by convolving a 
Gaussian random sequence with the square of the 
autocorrelation function of real data.} image 
of an ideal cluster model is transformed
to time stream data and passed through the full
iterative pipeline.  We use an isothermal beta
profile with $\beta=0.86$ and core radius 
$r_{r\rm c}=1^{\prime}$, typical parameters for
SZE clusters in \citet{marr11}.
Foreground and background SMGs are modeled 
by injecting 15 point sources with flux densities 
ranging between 5 and $20\,\rm mJy$ at random 
locations in the map. We define the transfer function
efficiency (TFE) as the ratio of flux densities of structures 
in the final map divided by  their ``true'' flux
densities in the input model image.
Point-source flux densities are measured using a 
least-squares fit to a circular Gaussian profile with a 
floating constant offset, and integrated SZE signal is measured as the
integrated flux density within a circular aperture of
radius $1.5^{\prime}$ (the typical radius used to extract 
SZE photometry in Section \ref{s-vpec}) after subtraction 
of point sources.  For a single-pass reduction, 
the TFEs of point sources and the extended SZE 
signal are \RawEfficiency{} and \RawSZEEfficiency{}, and increase
to \IterativeEfficiency{} and \IterativeSZEEfficiency{}
after running the full iterative pipeline.  The spatially 
extended SZE signal is not fully recovered because of
the pipeline's inability to faithfully recover signals on scales 
larger than $120^{\prime\prime}$.
Figure \ref{f-trans_image} displays the initial model, a 
single (non-iterative) 
reduction, and the final iterated image for our simulated
cluster, and Figure \ref{f-trans_phot} displays the input 
and output flux densities of individual point sources before
and after iteration.

For extended
SZE signals, the transfer function efficiency will
range between \RawSZEEfficiency{} and \IterativeSZEEfficiency{}
depending on whether the SZE signal was strong enough
to trigger the iterative filter.  The transfer function
efficiency for extended emission is ignored in the following
sections because identical filtering and aperture photometry
is used in all SZE wavebands, and we only require
relative photometry to derive peculiar velocities.
For SMGs, the flux that remains lost after iteration
is much lower than the typical statistical uncertainty of
our measurements and can be ignored.

\subsection{148\,GHz and 218\,GHz ACT}
We use the ACT\,148 and $218\,\rm GHz$
maps, with beam FWHMs of $1.4^{\prime}$ and $1.1^{\prime}$ 
\citep{hasselfield2013b}, respectively,  to constrain 
the low frequency portions of the clusters' SZE spectra. Before extracting the 
integrated flux densities
$S_{148}$ and $S_{218}$ for each cluster, we filter the
raw ACT maps to mimic the transfer 
function of our LABOCA observations, which
attenuates emission from spatial scales larger
than $\gtrsim 120^{\prime\prime}$.
This filtering is done by first taking the 
2D Fourier
transform of $1\,\rm deg^{2}$ cutouts
around the clusters in the ACT maps.  We
then apply the same frequency-domain 
filtering in 2D space for the ACT maps
that the BoA algorithm \texttt{flattenFreq}
performed on the 1D LABOCA time streams.  
To transform between temporal
and spatial frequencies, we use
the angular scanning speed of the LABOCA
observations.  We approximate the full continuous 35 
second APEX raster spiral as a collection of 
circular scans at many different radii.  One 
Fourier-filtered map 
is produced for each discrete scanning speed, and 
the final filtered maps used to extract
$148\,\rm GHz$ and $218\,\rm GHz$ 
flux densities are taken to be the
average of all individual velocity samplings.
Beam solid angles at 148 and $218\,\rm GHz$
are taken from \citet{hasselfield2013b}.

We have tested our Fourier-based replication 
of the LABOCA pipeline spatial filtering
by applying this Fourier filtering to the
ideal cluster model from Section 
\ref{ss-transfer}. We find that the fractional
SZE signal recovered within a $1.5^{\prime}$ 
aperture ($\rm TFE \simeq$\FourierSZEEfficiency{}) agrees with
the result from the real iterative pipeline 
($\rm TFE \simeq$\IterativeSZEEfficiency{}).

\subsection{2.1\,GHz ATCA}

\subsubsection{ATCA Observations}

We acquired $2.1\,\rm GHz$ Australia Telescope Compact 
Array (ATCA) observations of the ten clusters
in our sample with new LABOCA observations.
We used the $16\,\rm cm$-band receiver with the Compact Array 
Broadband Backend (CABB) giving 2049 $1\,\rm MHz$-wide 
channels spanning $1.1$--$3.1\,\rm GHz$.  
Observations were made in January 2011 (PI: Baker), 
December 2011 (PI: Baker), and April 2012 (PI: Lindner).  
All clusters have been observed with the 6A 
antenna configuration (baseline range  
$B=$628--$5939\,\rm m$), 
and ACT\,J0102-4915 has additional data in the 
1.5B configuration ($B=$31--$4301\,\rm m$).
For flux and bandpass calibration we used
PKS\,1934-638 \citep{reny94}.  
The Australia Telescope National Facility (ATNF)
on-line calibrator 
database\footnote{http://atoa.atnf.csiro.au}
was used to choose one or more suitably
bright, nearby, and compact phase calibrators
for each cluster; these are 
listed in Table \ref{t-obs-atca}.

\begin{deluxetable*}{llllc}
\tablecaption{ATCA $2.1\,\rm GHz$ observations}
\tablewidth{0pt}
\tablehead{
\colhead{} & 
\colhead{} & 
\colhead{} & 
\colhead{$t_{\rm obs}$\tnm{b}} &
\colhead{}
\\
\colhead{Month} & 
\colhead{Target} & 
\colhead{Phase cal} & 
\colhead{(hr)} & 
\colhead{Configuration}}
\startdata
Jan 2011  & ACT-CL\,J0232$-$5257         & J0214-522 &  19.8 & 6A \\
          & ACT-CL\,J0546$-$5345         & J0539-530 &  21.0 & 6A \\
Dec 2011  & ACT-CL\,J0102$-$4915         & J0047-579 &  12.1 & 6A \\
          & ACT-CL\,J0215$-$5212         & J0214-522 &  8.6  & 6A \\
          & ACT-CL\,J0235$-$5121         & J0214-522 &  8.5  & 6A \\
          & ACT-CL\,J0245$-$5302         & J0214-522 &  10.3 & 6A \\
          & ACT-CL\,J0330$-$5227         & J0334-546 &  8.8  & 6A \\
          & ACT-CL\,J0438$-$5419         & J0522-611 &  8.1  & 6A \\
          & ACT-CL\,J0559$-$5249         & J0539-530 &  8.9  & 6A \\
          & ACT-CL\,J0616$-$5227\tnm{a}  & J0539-530 &  7.8  & 6A \\
          &                            & J0522-611 &         & 6A \\
          &                            & J0647-475 &         & 6A \\
Apr 2012  & ACT-CL\,J0102$-$4915       & J0047-579 &  6.8    & 1.5B
\enddata
\tablenotetext{a}{Observations used three phase calibrators.}
\tablenotetext{b}{Total un-flagged, on-source integration time.}
\label{t-obs-atca}
\end{deluxetable*}

The software package MIRIAD \citep{saul95} was used
to calibrate, flag, invert, and clean the 
visibility data.
Radio frequency interference (RFI) that affected
all channels at a given time or at all times for a 
certain
channel was removed manually using \texttt{pgflag}
and \texttt{blflag}.
Transient RFI was removed using the 
automated flagging algorithm 
\texttt{mirflag} \citep{midd06}.
Baseline 1--2 in the April 2012 data 
contained powerful broad-spectrum RFI and 
was entirely flagged.
First-order multi-frequency synthesis images 
with robust parameter \texttt{robust}$=0$
were made using
\texttt{invert} and \texttt{mfclean}. 
Two rounds of self-calibration were then carried out, 
one solving for phase only, and
one for both phase and amplitude together.
The final RMS sensitivities at phase center
vary over the range 6.9--$12\,\rm \mu Jy\,beam^{-1}$
(see Table \ref{t-atca-maps}).

\begin{deluxetable*}{cccccc}
\tablecaption{ATCA map properties}
\tablewidth{0pt}
\tablehead{
\colhead{} & 
\colhead{$\left< \nu \right>$\tnm{a}} & 
\colhead{$\theta_{\rm major}$\tnm{b}} & 
\colhead{$\theta_{\rm minor}$\tnm{c}} & 
\colhead{P.A.\tnm{d}} & 
\colhead{map RMS\tnm{e}} 
\\
\colhead{Target} &
\colhead{(GHz)} &
\colhead{($\prime\prime$)} &
\colhead{($\prime\prime$)} &
\colhead{($\circ$)} &
\colhead{($\mu \rm Jy\, beam^{-1}$)}}
\startdata
ACT-CL\,J0102$-$4915 & 2.15 & 6.13 & 3.09 &  -1.9 &  7.5 \\
ACT-CL\,J0215$-$5212 & 2.16 & 4.66 & 2.93 & -21.0 & 11.0 \\
ACT-CL\,J0232$-$5257 & 2.13 & 4.80 & 3.06 &   6.0 &  8.1 \\
ACT-CL\,J0235$-$5121 & 2.17 & 5.26 & 2.71 & -10.0 & 10.9 \\
ACT-CL\,J0245$-$5302 & 2.16 & 4.44 & 2.97 &   4.3 & 10.5 \\
ACT-CL\,J0330$-$5227 & 2.17 & 5.06 & 2.73 &  17.7 & 11.7 \\
ACT-CL\,J0438$-$5419 & 2.15 & 5.30 & 2.80 & -19.6 & 11.9 \\
ACT-CL\,J0546$-$5345 & 2.13 & 4.66 & 3.21 &  -3.1 & 6.9  \\
ACT-CL\,J0559$-$5249 & 2.15 & 5.35 & 2.86 & -13.1 & 9.6  \\
ACT-CL\,J0616$-$5227 & 2.15 & 4.99 & 2.98 &  26.2 & 12.0
\enddata
\tnt{a}{Effective frequency (different for different clusters due to RFI flagging)}
\tnt{b}{Synthesized beam major axis}
\tnt{c}{Synthesized beam minor axis}
\tnt{d}{Synthesized beam position angle}
\tnt{e}{RMS noise at the phase center}
\label{t-atca-maps}
\end{deluxetable*}

\subsection{$Herschel \; observations$}
To help constrain the high-frequency end of the SZE spectrum when
deriving cluster peculiar velocities 
(see Section \ref{s-vpec} below), we use newly obtained 
250, 350, and $500\,{\rm \mu m}$ imaging of six targets
in our sample using the Spectral and Photometric Receiver 
\citep[SPIRE; ][]{griffin2010} instrument on board the {\it Herschel Space
Observatory} \citep{pilb10}.\footnote{{\it Herschel}
proposal ID = OT2\_abaker01\_2}  The six targets were prioritized
within our ATCA sample of ten on evidence of having high mass
(SZE decrement strength, optical richness, etc. from
\citet{mena10} and \citet{sifo13}), although in
the end they did not overlap perfectly with the five objects 
for which we detected strong SZE increments with
LABOCA.  All SPIRE observations used the Large Map mode with
nominal map speed.  For five of the six clusters observed
(ACT-CL\,J0102-4915, ACT-CL\,J0235-5121, ACT-CL\,J0245-5302,
ACT-CL\,J0438-5419, and ACT-CL\,J0546-5345), four repetitions
apiece of $8^\prime \times 8^\prime$ maps centered at four
different dither positions yielded a total of $4 \times 4 \times
139\,\rm s = 2224\,{\rm s}$ of time on-source.  For ACT-CL\,J0330-5227,
we were able to fit in four repetitions of a $6^{\prime} \times 6^{\prime}$
map at a single position, for a total of $1 \times 4 \times 123\,\rm s
 = 492\,{\rm s}$ on-source, and, consequently, a much shallower
depth.  All data were reduced using the {\it Herschel} Interactive
Processing Environment \citep[HIPE, version 10.3.0; ][]{ott2010} using standard scripts.
The final maps have angular resolutions 17.6, 23.9, and
$35.0^{\prime\prime}$ with mean confusion-limited RMS map 
sensitivities of 7.4, 7.2, $7.2\,{\rm mJy\,beam^{-1}}$ in 
the 250, 350, and $500\,{\rm \mu
m}$ bands, respectively, for all but ACT-CL\,J0330-5227.

The typical increment intensities {\it relative} to those at
345\,GHz are expected to be approximately 0.05\%, 2\%, 
and 34\% in observations at 250, 350, and $500\,{\rm \mu
m}$, respectively.  The high sensitivity and comparatively low
angular resolution of {\it Herschel}/SPIRE, however, mean that
our images all suffer badly from confusion: their brightness
fluctuations are primarily due to the blending together of bright,
unresolved sources (here, SMGs).  Section 3.6 describes the
approach we have used to disentangle SZE increment signal
from this confused background.

\subsection{Foreground Galactic dust subtraction}
The SPIRE maps of ACT-CL\,J0245-5302 and ACT-CL\,J0546-5345
contain detectable diffuse emission from Galactic dust on large spatial
scales.  The mean dust temperatures within $10^{\prime}$ radii 
(covering the full extent of the SPIRE maps) of the cluster
centers as measured by {\it Planck} are
$T^{0245}_{D} = 19.9 \pm 0.7\,\rm K$ and
$T^{0546}_{D} = 20.1 \pm 0.3\,\rm K$ \citep{abergel2013}.
To remove the foreground dust emission from 
the maps, 
we first generate a model of the smoothly varying foreground 
dust signal by applying a median filter with kernel 
size $4^{\prime}$ to the $250\,\rm \mu m$ maps.  This
dust model image is then jointly fit to similarly filtered
versions of the 350 and $500\,\rm\mu m$ maps using a 
modified blackbody thermal spectrum with emissivity 
fixed at $\beta=1.57$, the value measured in the directions of
both clusters by \citet{abergel2013}.
The $250\,\rm\mu m$ dust model is then scaled according
to the best-fit dust temperature and subtracted from 
the $350\,\rm\mu m$ and $500\,\rm\mu m$ maps.

We find best-fitting dust temperatures of
$T^{0245}_{D} = 23_{-1}^{+2}\,\rm K$
and $T^{0546}_{D} = 19_{-1}^{+1}\,\rm K$.  The 
reasonable consistency of our temperatures with
the {\it Planck} results supports
our assumption that Galactic dust is the cause of the 
observed diffuse signals.

\subsection{The Sunyaev Zel'dovich effect}

The deflection in CMB intensity
due to the thermal SZE for a 
single-temperature gas is \citep{zeld69,sunyaev1970}
\begin{equation}
  \label{e-tSZ}
  \Delta I_{\rm tSZ} =y \, I_{0} \, g(x,T_{e}), 
\end{equation}
in terms of the scaled frequency 
$x\equiv h\nu/k_{\rm B}T_{\rm CMB}$, 
$I_{0}=2(k_{\rm B}T_{\rm CMB})^3/(hc)^2=22.87\,\rm Jy\,arcmin^{-2}$, 
the electron temperature $T_{e}$, and the Compton 
parameter $y$ (see Equation \ref{e-compton}).
The function $g\left( x, T_{e} \right)$ 
is the derivative of the Planck function ($dB_{\nu}/dT$)
multiplied by the SZE spectrum,
\begin{equation}
    g(x,T_{e}) = 
    \frac{x^4 e^x}{\left(e^x-1\right)^2} \left( x\frac{e^x+1}{e^x-1} -4\right) 
    \left[ 1+\delta_{\rm tSZ}\left( x,T_{e}\right)\right], 
    \end{equation}
where $\delta_{\rm tSZ}$ represents relativistic 
corrections that become important at high electron 
temperatures, especially when $x\gg1$.

The change in CMB intensity due to the 
kSZ effect is given by \citep{sunyaev1972}
\begin{align}
\label{e-kSZ}
    \Delta I_{\rm kSZ}& =-\tau_e\,I_0\,\left(\frac{v_p}{c}\right)
    \frac{dB_\nu}{dT} \nonumber \\
     & = -\tau_{e}\,I_0\,\left(\frac{v_{p}}{c}\right)
\frac{x^4 e^x}{\left(e^x-1\right)^2} 
\left[ 1+\delta_{\rm kSZ}\left( x,T_{e}\right)\right],
\end{align}
where $v_p$ is the cluster line-of-sight 
peculiar velocity with respect to the CMB in the
cluster rest frame,  the optical depth to electron
scattering $\tau_e=\int \sigma_{\rm T}\,n_e\, dl$, 
and $\delta_{\rm kSZ}$ represents 
higher-order relativistic corrections.

Equations \ref{e-tSZ} and \ref{e-kSZ} are shown
to highlight the leading order terms in the
thermal and kinetic SZE signal. In our analysis,
we do not use the analytic expressions, but instead
compute the SZE signal with numerical integration 
using the \texttt{C++} package 
\texttt{SZ{\footnotesize PACK}} \citep{chlu12,chlu13}.
SZpack allows for quick computations of the relativistic
tSZ and kSZ signals with $10^{-5}$ precision for 
temperatures up to $T_e\simeq 25\,\rm keV$ at 
frequencies up to and including the high-frequency
{\em Herschel} SPIRE bands ($x\lesssim 20$) and
peculiar velocities up to $|v_{\rm pec}|/c\simeq 0.1$ 
($|v_{\rm pec}|\simeq 30\times 10^{3}\,\rm km\,s^{-1}$).

\subsection{Isolating SZE signal in SPIRE maps}
\label{ss-spire_pipeline}

We isolate the SZE signal in the $350\,\rm \mu m$ and 
$500\,\rm \mu m$ SPIRE maps by using the $250\rm\,\mu m$ 
map (which contains a negligible contribution from SZE signals
compared to the 350 and $500\,\rm\mu m$ bands) 
to derive a model for the confused SMG background.
This approach was presented by \citet{zemc10}, but 
our implementation of
it differs in several important respects.
For a given cluster, we begin by producing a model 
of the background SMG 
confusion signal at $250\,\rm \mu m$ ($\mathcal B_{250}$) 
by applying a median filter with kernel size 
$\rm 3\times\,FWHM_{250}$ 
to the $250\rm\,\mu m$ image.
We then smooth $\mathcal B_{250}$ to the angular resolution
of the $350\,\rm \mu m$  image to
produce an estimate of the confused background at 
$350\,\rm \mu m$, $\mathcal B_{350}$.

We next locate bright 
sources that stand out from the confused 
$250\rm \,\mu m$ background by first subtracting 
$\mathcal B_{250}$ from the original $250\rm \,\mu m$ 
image, and then searching for point sources in the resulting
background-subtracted image using a spatial 
matched filter \citep[e.g., ][]{serj03}. The matched
filter identifies high S/N peaks, 
which are then individually fit to an ideal 
$250\,\rm \mu m$ PSF with positions and 
flux densities allowed to vary.    Sources are extracted 
to a S/N level of $4.5\sigma$;  
we find 100--200 sources per cluster.
The properties of the  $250\,\rm \mu m$ point sources 
represent a starting point for modelling bright SMGs at $350\,\rm \mu m$.

The $350\,\rm \mu m$ map is then fit to a 
$3N + 2$-parameter model consisting of the confused background
plus bright SMGs.  The offset $a$ and scaling $b$ of
the confused background template and the remaining $3N$ parameters
representing the coordinates $\vec{x}^{\prime}$ and flux densities 
($S$) of bright SMGs are combined into a
parameter vector 
$\theta = ( \{\vec{x}^{\prime}_i,S_i\}_{n=1}^{N}, a, b)$.
The resulting model
\begin{equation}
    {\mathcal M}(\vec{x};\theta) =  
    a + b\,{\mathcal B_{350} + \sum_n^N S_n \, e^{-\left| \vec{x}-\vec{x}^{\prime}_n\right|^2/2\sigma_n^2}
     },
\end{equation}
is then fit to the $350\,\rm \mu m$ image using 
least-squares minimization.  The optimization is carried out using 
Powell's method \citep{powell1964}, where in each iteration the 
parameters are minimized
individually.  Powell's method works well for this problem because 
each source affects only a small fraction of the data.
This joint optimization naturally accommodates the increased 
blending that occurs in the longer
wavelength images.
A $350\,\rm \mu m$ source is considered an SMG and subtracted from
the map if its final converged position is less than
a distance $0.25\,\rm FWHM_{350}$ 
from the location for that source intially guessed from 
the $250\,\rm\mu$ data.
To avoid fitting the confusion model to
the cluster signal itself, the above optimization is carried 
out once while masking in central cluster region and allowing all 
parameters in $\theta$ to vary, then
once more without the mask and holding $a$ and $b$ fixed
at their previously fit values.

The process described above is 
repeated (independently, but in the same way) 
for the SPIRE $500\,\rm \mu m$ map.
Emission that remains in the $350\,\rm \mu m$
and $500\,\rm \mu m$ maps after subtraction of the
$250\,\rm \mu m$-inspired SMG background model is due
either to (a) SZE increment, or (b) SMGs that were not
detected in the $250\,\rm \mu m$ map at all, possibly
because they lie at very high redshifts.
After removing the SMG confusion noise, the 
350 and $500\,{\rm \mu m}$ maps have mean
RMS sensitivities of 2.9 and 
$2.7\,{\rm mJy\,beam^{-1}}$,  respectively.

\begin{figure}
    \centering
    \includegraphics[scale=0.85]{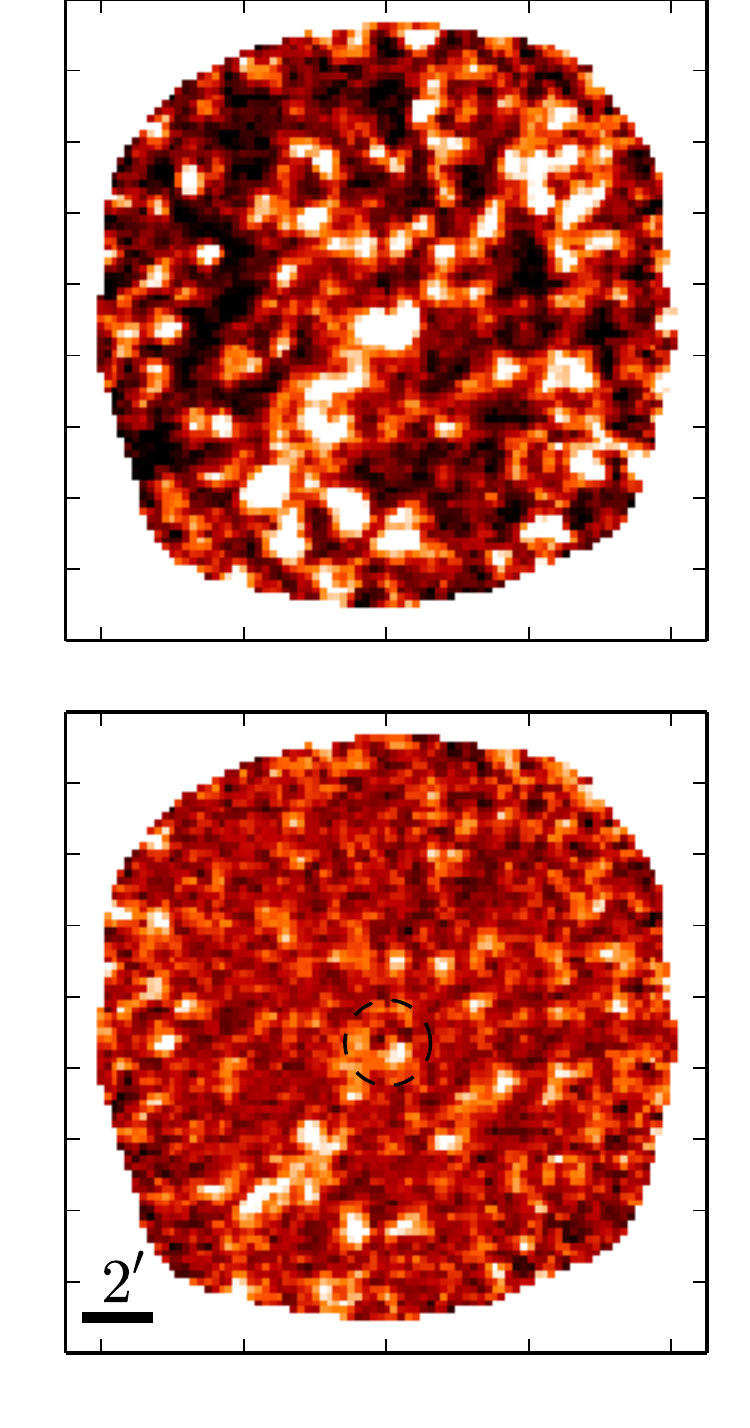}
    \caption{Confusion signal removal process for
    cluster ACT--CL\,J0102-4915.  The upper panel
    shows the original SPIRE $500\,\rm \mu m$ image
    with confusion-dominated RMS noise of 
    $6.8\,\rm mJy\,beam^{-1}$.  The color scale ranges linearly
    between -10 and $10\,\rm mJy\,beam^{-1}$.
    The lower panel shows
    the same map after running the SZE-extraction from
    Section \ref{ss-spire_pipeline} to remove the confused
    SMG background; here the RMS noise is reduced to $2.2\,\rm mJy\,beam^{-1}$.
    The black
    dashed circle is centered on the ACT\,148\,GHz centroid 
    (J2000: 01:02:53, -49:15:19) 
    and has radius of $1^{\prime}$.}
    \label{f-SPIRE_3panel}
\end{figure}
\begin{figure}
    \includegraphics[scale=0.45]{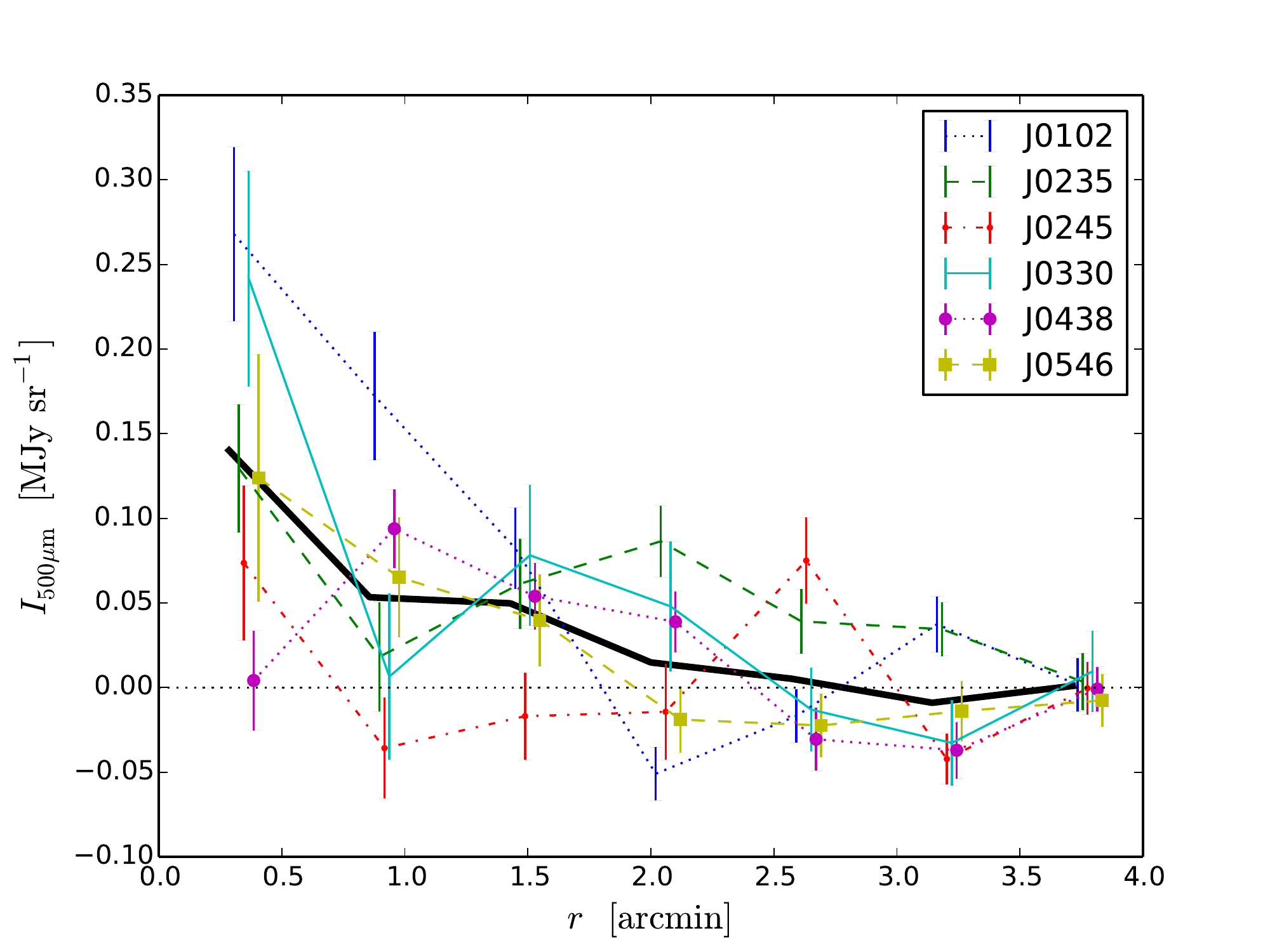}
    \caption{$500\,\rm\mu m$ SZE radial profiles for
             clusters with {\em Herschel} SPIRE coverage.
             Profiles are extracted from maps after
             isolating the SZE signal component 
             (Section \ref{ss-spire_pipeline}), Fourier filtering
             the resulting map to mimic the LABOCA transfer
             function, and extracting point-like sources using
             the source extraction algorithm from Section 
             \ref{ss-smgs}.  The thick black line represents
             the mean profile.\\}
\label{f-profiles}
\end{figure}

Figure \ref{f-SPIRE_3panel} shows an example of
the $500\,\rm \mu m$ SZE extraction process in
ACT--CL\,J0102-4915.  The upper panel shows the
original SPIRE $500\,\rm \mu m$ image.  The
lower panel shows the image after 
removal of the confused 
SZ background using the technique described above
(and after Fourier filtering the image to simulate
the LABOCA transfer function).  Radial profiles of 
the resulting SZ signals for clusters 
with {\em Herschel} SPIRE coverage are shown 
in  Figure \ref{f-profiles}.

To test the reliability of our data reduction pipeline and
source extraction algorithm, we compared our 
$250\,\rm\mu m$ integral number counts to those from the
{\em Herschel} Multi-tiered Extragalactic Survey 
\citep[HerMES; ][]{oliver2010}.  Because our point-source
removal technique breaks up $350$ and $500\,\rm\mu m$ 
sources and divides their flux among multiple $250\,\rm\mu m$ 
counterparts, our number counts for $350$ and $500\,\rm\mu m$
cannot be meaningfully compared to those of
\citet{oliver2010}.  
We can however compute the number 
density of sources with $S_{250}>100\,\rm mJy$ around all
clusters with {\em Herschel} SPIRE data except 
ACT-CL\,J0330-5227, which has significantly reduced sensitivity,
also masking within $5^{\prime}$ radii of the cluster centers to avoid
the effects of gravitational lensing.  
At these bright flux densities where completeness corrections
will be minimal, we find
$N_{>100\,\rm mJy} = 8.0 \pm 4.0 \,\rm deg^{-2}$, in agreement with
the results from \citet{oliver2010} who find 
$N_{>100\,\rm mJy}=12.8 \pm 3.5\,\rm deg^{-2}$.

\section{Point source contamination}
\label{s-contam}

\subsection{SMGs}
\label{ss-smgs}
In the iteratively reduced LABOCA images, we find bright SMGs 
superposed on the clusters' extended SZE emission.  A complete 
analysis of the statistical properties of the detected  
SMGs and their multi-wavelength counterparts will be presented 
in a forthcoming paper (Aguirre et al., in prep).  Here 
we only consider the SMGs 
nearest to the clusters and study how their presence affects 
the measured SZE signals.

SMGs are extracted from the iterative multi-scale 
LABOCA map (Section \ref{s-iterative}) of a given cluster using 
the following process.  First, SMG locations are found 
by using a spatial matched filter \citep{serj03} to search
for sources shaped like the ideal telescope PSF within an
unsharp-masked version of the map.  The unsharp mask
uses a median filter with a kernel size of 
$3 \times \rm FWHM$, and allows for point sources
to be disentangled from the diffuse large spatial-scale 
SZE signals near the cluster centers. 
Next, we measure the flux density of each detected SMG 
by fitting the original (non un-sharp masked) data 
at the source's location with an ideal PSF while
holding the position fixed and varying the flux density 
and a non-negative background offset.

We detect a total of 
\NoOfSMGInRFH{} $\rm S/N>4$ point sources 
within $\theta_{500c}$ of our 11 cluster centers,
giving an estimated total flux density per cluster
in high-significance SMGs of \LabocaContamRFH{}.
Using the combined area of \TotalSmgSolidAngle{},
this sample has cumulative number counts of
$>10\,\rm mJy$ SMGs of \SmgCounts{}, 
6--10 times greater than those of
sources with comparable brightness in 
blank fields \citep{weiss2009}.
Due to the negative k-corrections and high redshifts
of SMGs, the number density enhancement is
likely due to gravitational lensing by the clusters' 
potentials.
Gravitational lensing 
does not alter the average integrated flux density of
the projected SMG population, but it does increase the
Poisson ``shot noise'' 
variance \citep[e.g., ][]{refregier1997}, and could introduce
a bias in flux-constrained (i.e., either a S/N threshold or
upper limit) aperture photometry
near vs. away from clusters.

The $345\,\rm GHz$ flux density in bright SMGs 
inside the smaller apertures used in our analysis
of the SZE (Section \ref{s-vpec}) is \SmgContamLaboca{}, 
corresponding to \SmgContamArTwo{} 
at $218\,\rm GHz$ and \SmgContamArOne{} at $148\,\rm GHz$.
The scaling to lower frequencies is done using a modified 
Planck function with parameters $\beta=1.5$ and $T_{\rm D}=37.4\,\rm K$, 
 values for typical blank-field 
$345\,\rm GHz$-selected SMGs \citep{ward11}. The contaminating 
SMG signal represents \SmgContamLabocaFrac{} of the SZE 
signal per cluster at $345\,\rm GHz$, and \SmgContamArOneFrac{} at 
$148\,\rm GHz$, respectively.  
The contaminating SMGs are subtracted from the maps before 
further analysis of the SZE signals.

\subsubsection{El Gordo and ``La Flaca''}

\begin{figure}
\centering
\includegraphics[scale=0.45]{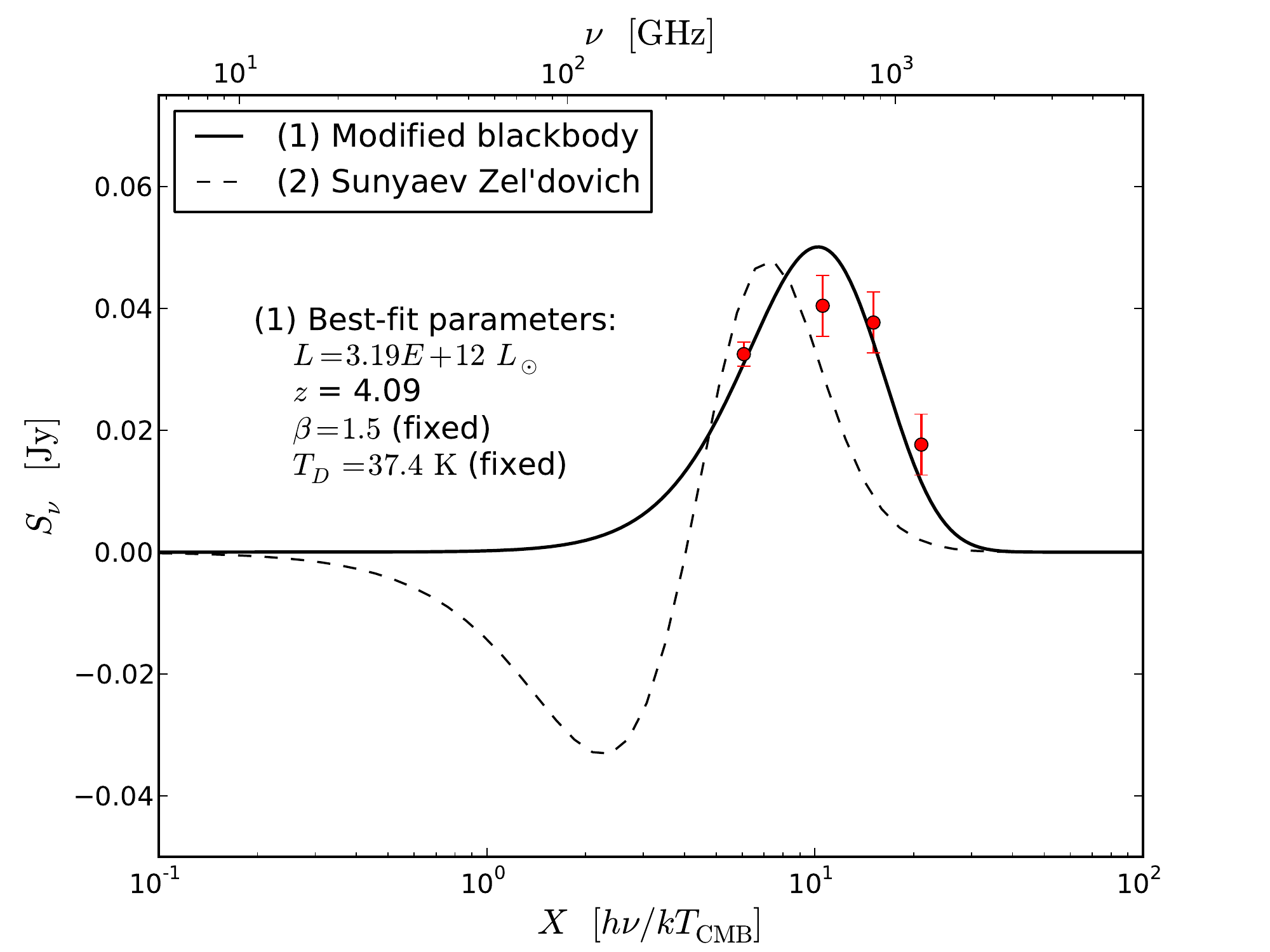}
\caption[Spectral energy distribution for ``La Flaca'']
{Observed photometry and best-fit thermal and SZE
spectra for J010256$-$491509 (``La Flaca'').  From left
to right, the red points represent LABOCA\,$345\,\rm GHz$ 
and SPIRE\,$500\,\rm\mu m$, $350\,\rm\mu m$, 
and $250\,\rm\mu m$ data.  The solid black line represents the 
least-squares fit to the data using a modified 
blackbody spectrum with fixed parameters $T_{d}=37.4\,\rm K$ 
and $\beta=1.5$, which are adopted from \citet{ward11}. The best
fit redshift and infrared luminosity are \LaflaccaRedshift{} and 
\LaflaccaLIR{}.  The dashed curve
represents the best-fitting thermal SZE SED 
assuming $T_{e}=25\,\rm keV$, which is formally
unsatisfactory.}
\label{f-laflacca}
\end{figure}

El Gordo (ACT-CL\,J0102-4915) has two bright point sources
projected at small clustercentric radii; the brighter
of the two ($S_{\nu}=36\pm 2\,\rm mJy$)  we refer
to as ``La Flaca.''  We briefy describe La Flaca's
properties here because it is exceptionally bright
(The next brightest SMG within our sample has flux 
density $\sim 27\,\rm mJy$ and is located in 
ACT--CL\,J0438-5419), second in  brightness
only to the lensed SMG \citep[e.g.]{wilson2008,joha10} 
behind the Bullet cluster (ACT-CL\,J0658-5557).
La Flaca is located at $\alpha=15.732207^{\circ}$ 
$\delta=-49.252415^{\circ}$ and has an
extremely faint radio counterpart 
with $S_{2.1}= 24\pm 7\,\rm\mu Jy$.
To test the possibility that La Flaca is
a local SZE enhancement, we compare the spectral fits
of the source's LABOCA+SPIRE photometry using 
a modified black body spectrum and thermal SZE
spectrum.  Figure \ref{f-laflacca} shows the
best-fit curves for each model.  We find that
the SZE spectral shape is incompatible with
the data; thus, La Flaca is 
more likely to be
a high-redshift dusty star-forming 
galaxy than a compact SZE enhancement.

The best-fit modified Planck function
parameters are $z=4.1$ and 
$L_{\rm IR}=3.2\times 10^{12}\,L_{\odot}$, 
assuming the median observed 
$T_{\rm D}=37.4\,K$ and $\beta=1.5$ from the
LABOCA Extended-CDFS SMG survey
\citep[LESS; ][]{ward11}.
Using the $610\,\rm MHz$ GMRT data 
\citep{lindner2014}, we derive a radio
spectral index of $\alpha\sim -1.3$, giving a
redshift estimate of $z=4.0$ based on the 
radio-to-far IR spectral index \citep{cari99}, 
in agreement with
that of the spectral fit and further supporting the
conclusion that La Flaca is a high-redshift SMG.

There are two faint HST sources within 
$1^{\prime\prime}$ of La Flaca's 345\,GHz centroid.
The brighter source has magnitude $m_{775}=23.8$
and resembles a spiral galaxy, and the fainter source
has magnitude $m_{775}=26.8$ and is unresolved.
La Flaca is also only a few arcseconds away
from the positions of gravitational lensing 
critical curves for background galaxies 
between $z=4$--9 \citep{zitrin2013}.

\subsection{Radio sources}

Radio and IR-bright sources 
can potentially ``fill in'' SZE decrement signals
and artificially enhance SZE increment signals in the
$1.5^{\prime}$-resolution SZE maps typically used
to detect clusters.  
With our high-resolution $2.1\,\rm GHz$
and $345\,\rm GHz$ imaging, we can disentangle
the signals of point sources from those of
the true SZE signal, allowing us to 
derive more robust 
measurements of the thermal and kinetic 
SZE signals. In this section, we quantify the
degrees of radio and submillimeter 
contamination of the SZE signal.

\subsubsection{$2.1\,\rm GHz$ number counts}

We use the 
Common Astronomy Software Applications 
\citep[CASA; ][]{mcmullin2007} task
\texttt{findsources}
to fit elliptical Gaussian profiles to all bright sources
in the ATCA maps down to a significance of
$4\sigma$ and out to the $10\%$ primary 
beam power radius ($22^{\prime}$). 
We adopt a 
primary beam profile based on the 
effective mean frequency of each 
observation (see Table \ref{t-atca-maps}).
To minimize radial selection bias, 
we only consider sources that
have primary beam-corrected flux 
densities above the
detection threshold at {\em all} radii,
i.e., $>4\sigma/0.1=400\,\rm\mu Jy$. 
Additionally, only compact sources that have major and 
minor FWHMs, $a$ and $b$, satisfying 
$a < 3 \rm\times  FWHM^{\rm beam}_{\rm maj}$, 
$ b > 0.5 \times \rm FWHM^{\rm beam}_{\rm min}$, 
and $a / b < 1.5 \times \rm FWHM^{\rm beam}_{\rm maj}/FWHM^{\rm beam}_{\rm min}$ are kept.
We consider only compact sources because extended radio 
sources like jets and lobes from active galactic nuclei have 
steep radio spectral indices and will therefore 
contribute negligibly in the frequency 
range $\nu \ge 148\,\rm GHz$.
These criteria
result in a total for all clusters of 
\NoOfAtcaSources{} radio sources with flux
densities between \BinLow{} 
and \BinHigh{}.

Fainter sources have an increased chance of 
having their flux densities scattered below
the detection threshold by noise, thus lowering
their completeness.  
Because the noise in the $2.1\,\rm GHz$ maps
is nearly Gaussian, we correct for completeness
using an analytic correction in the following way.   
The completeness probability, $C$, that a 
source with true flux density $S_{\nu}$ will be 
detected above
a threshold of $N\,\sigma$ is given by
\begin{equation}
C=1-\Phi\left(\frac{N\sigma-S_{\nu}}{\sigma}\right),
\end{equation}
where $\Phi(x)$ is the cumulative normal distribution
function.  The completeness correction is then
implemented by computing number counts using
a weighted histogram with weights $w_i=C_i^{-1}$.

\begin{figure}
\centering
\includegraphics[scale=0.50]{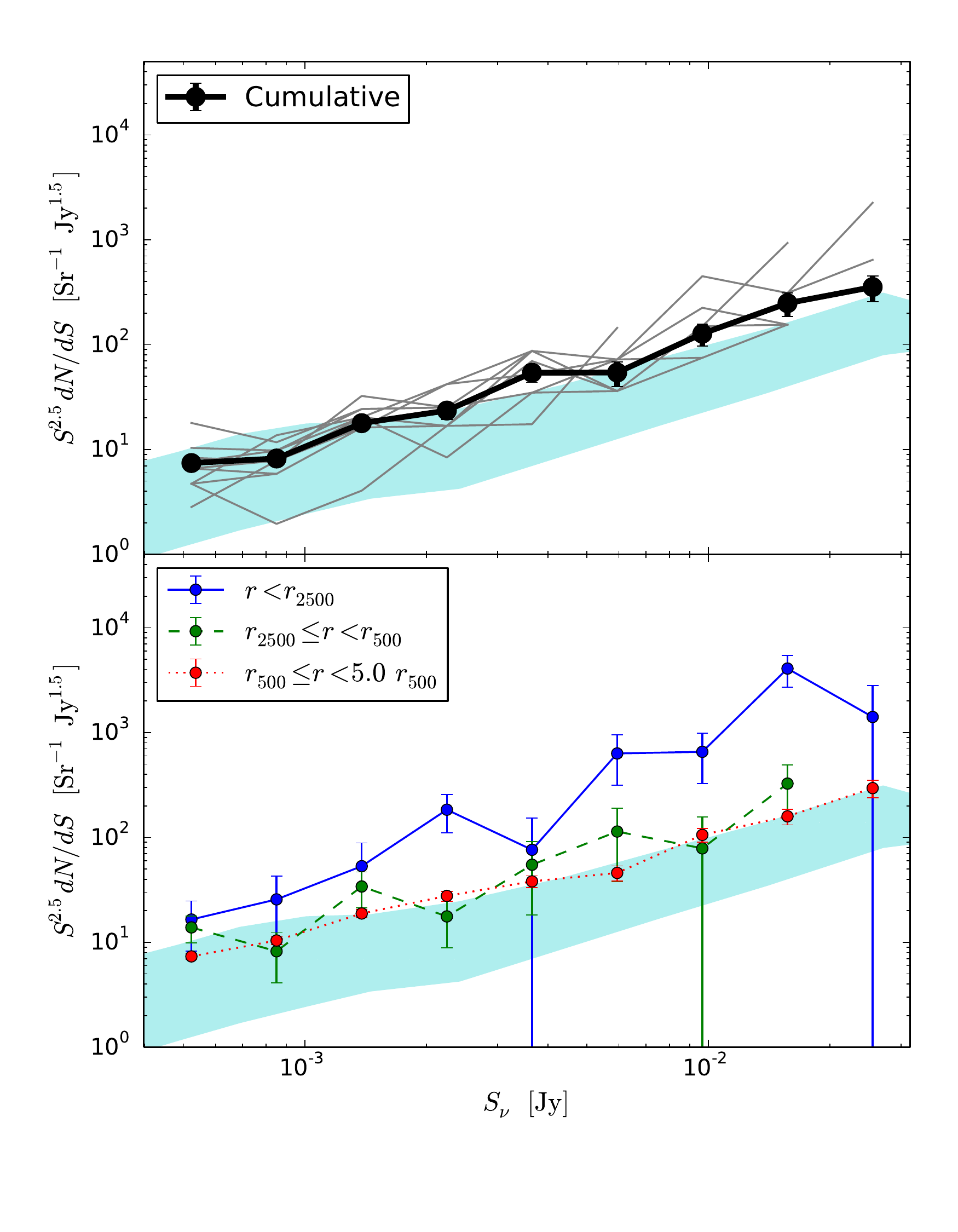}
\caption[$2.1\,\rm GHz$ number counts]
{ $2.1\,\rm GHz$ number counts. {\em Upper panel} 
Total $2.1\,\rm GHz$ differential number counts 
towards the clusters out to a maximum search radius of
\MaximumRadius{}.  The grey lines
represent the counts around individual clusters.  The 
thick black line represents the cumulative counts across
all clusters.  The black circles mark the centers of the
bins, which are logarithmically spaced 
between $400\,\rm \mu Jy$ and $31\,\rm mJy$.
{\em Lower panel: }
$2.1\,\rm GHz$ differential number counts 
as a function of physical clustercentric
projected radius.  The connected points 
and error bars represent number 
counts and Poisson uncertainties inside a circle with projected 
radius $r \leq r_{2500c}$ (solid), inside
an inner annulus with radii $r_{2500c} < r \leq r_{500c}$ (dashed), and
inside an outer annulus with radii $r_{500c}< r <  5.0\,r_{500c}$
(dotted).  In all panels, the shaded blue area shows the
$95\%$ confidence region of ATCA $20\,\rm cm$ number counts in
the AKARI Deep Field South from \citet{white2012}.  We use the
conversion $r_{2500c} \simeq 0.44\,r_{500c}$ \citep{arna10}.}
\label{f-counts-all}
\end{figure}

Figure \ref{f-counts-all} (upper panel) shows combined
and individual $2.1\,\rm GHz$ number counts 
in nine log-spaced flux density bins. 
The power-law index of the counts $\delta$, 
where $dN/dS\propto S^{-\delta}$, is
$\delta\simeq 1.7$.  
We next compute the number counts in three
disjoint regions defined relative to the clusters' physical
radii, and find that the counts in the 
$\theta < \theta_{2500c}$ region are elevated (especially at
high flux densities) compared to the more distant 
regions (Figure \ref{f-counts-all} lower panel), in agreement
with a previous radio survey of southern X-ray selected 
clusters \citep{slee2008}.  This enhanced
source density in cluster interiors is likely 
due to the presence of both
radio-loud cluster member galaxies
and (possibly at a lower level) 
gravitationally lensed background AGNs and
star-forming galaxies.
We have also checked for variation in the counts 
for the top, bottom, left, and right halves 
of the radio images and find no significant
differences.

\subsubsection{Radio contamination extrapolated to $148\,\rm GHz$}

\begin{figure}
\centering
\includegraphics[scale=0.6]{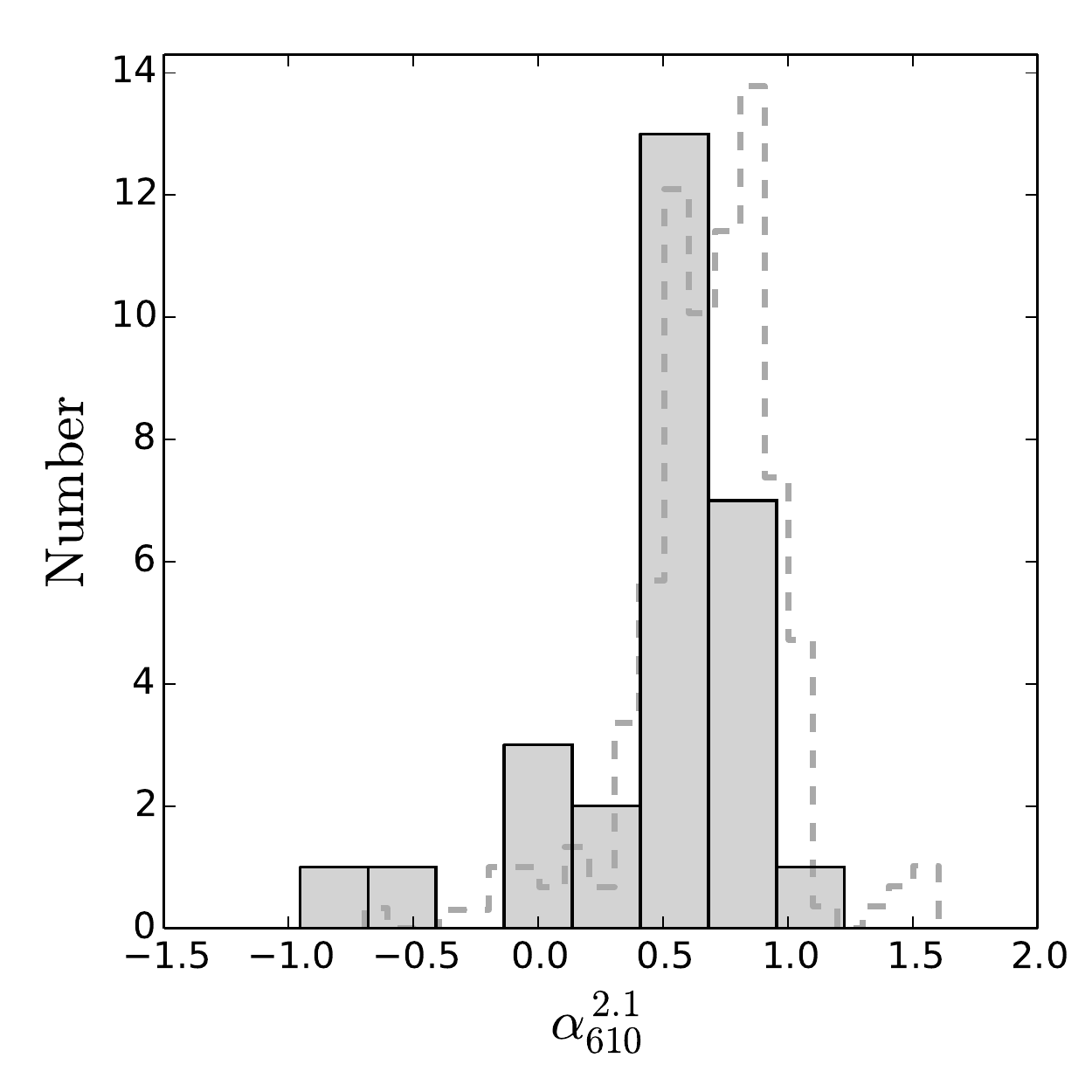}
\caption[Radio source spectral index distribution]
{Spectral index $\alpha^{2.1}_{610}$
of \NoOfATCAGMRTPairs{} pairs of $>4\sigma$ significance 
$2.1\,\rm GHz$ and $610\,\rm MHz$ sources within \AlphaSearchRadius{}
of the center of ACT-CL\,J0102$-$4915.  The median and interquartile 
range are \AlphaMedian{} and \AlphaInnerQuartile{}, respectively.  
For comparison, the grey dashed line shows the 
$\rm 1.4\,\rm GHz/330\,\rm MHz$ spectral index distribution of 
radio sources detected from deep observations of 
the Lockman Hole North \citep{owen09}.}
\label{f-spectral_index}
\end{figure}

We use our $610\,\rm MHz$ Giant Metre-wave Radio
Telescope (GMRT) image of ACT-CL\,J0102$-$4915 
\citep{lindner2014} to compute
the radio spectral index 
$\alpha^{2.1}_{610}$ distribution
of $2.1\,\rm GHz$ sources, which allows us
to
extrapolate their flux densities
to the higher frequencies of the SZE decrement.
We find \NoOfATCAGMRTPairs{}  
$2.1\,\rm GHz$/$610\,\rm MHz$ pairs of
sources with $ <1^{\prime\prime}$ ($7.8\,\rm kpc$) separation 
and $>4\sigma$ significance out to a $6^{\prime}$ radius 
relative to the SZE decrement centroid,
and compute their spectral indices as
\begin{equation} 
  \alpha^{2.1}_{610}=-\log(S_{2.1}/S_{610})/\log(2.1/0.610).
\end{equation}
The distribution $\alpha^{2.1}_{610}$ is shown
in Figure \ref{f-spectral_index}.  
We find a median spectral index
\AlphaMedian{} with interquartile range 
\AlphaInnerQuartile{}.
Only $4.9\%$ of $>4\sigma$ 
ATCA radio sources have
optically-selected cluster 
member counterparts (for the clusters
that overlap between the two samples) from \citet{sifo13}.  The
disjoint source catalogs along with the fact that our 
spectral index distribution is similar to that of 
\citet{owen09} indicates that radio source contamination 
in our clusters is due mostly to foreground and 
background galaxies, allowing us to use the spectral 
index distribution measured in the field of 
ACT-CL\,J0102-4915 to predict the 148\,GHz 
contamination in our full cluster sample.
The recent study of the average submillimeter emission
associated with 1.4\,GHz-selected AGN by \citet{gralla2014}
finds that a low-frequency estimate of average AGN spectral
index remains valid when extrapolating all the
way to $218\,\rm GHz$, validating our extrapolation of
2.1\,GHz emission up to the ACT bands at 148 and 218\,GHz.

Because we only measure the spectral indices of
sources that are detected at both $610\,\rm MHz$
and $2.1\,\rm GHz$, we check for sources with 
flat/inverted spectral
indices that may not be detected at $610\,\rm MHz$
but could be bright at $2.1\,\rm GHz$ and higher
frequencies by searching for $20\,\rm GHz$ counterparts
to our ATCA and GMRT detections using the
Australia Telescope $20\,\rm GHz$ survey \citep[AT20G; ][]{murphy2010}.
There is one match between the AT20G catalog
\citep{murphy2010} and our ATCA and GMRT sources.
The source lies $15^{\prime}$ away from the center
of cluster ACT-CL\,J0215-5212 and 
has $S_{20}=0.11\pm 0.01 \,\rm Jy$
and $S_{2.1} = 0.011867 \pm 0.001726 \,\rm Jy$, giving
an inverted spectral index of $\alpha=1.0$.
This source is only $10^{\prime\prime}$ away
from a much brighter ATCA source with 
$S_{2.1} = 0.897 \pm 0.001\,\rm Jy$.

We detect \NoOfRadioInRFH{}
$2.1\,\rm GHz$ sources with $S_{2.1}>4\sigma$ 
within $\theta_{500c}$ of our ten clusters 
with ATCA imaging, corresponding to
a mean $2.1\,\rm GHz$ flux density within $\theta_{500c}$
per cluster of \AtcaContamRFH{}.
After scaling the flux density to
the $148\,\rm GHz$ SZE decrement using 
\AlphaMedian{} and including an additional
uncertainty based on the 
variation in
the spectral index between
the extremes of the 
interquartile range, the typical contaminating
radio signal from $2.1\,\rm GHz$-detected 
synchrotron radio sources is estimated to lie 
between \AtcaContamArOneRFH{} in the 
$148\,\rm GHz$ SZE decrement.

Inside the photometric apertures
used to constrain the clusters' peculiar 
velocities (Section \ref{s-vpec}), 
radio sources contribute \RadioContamRadio{},
corresponding to 
\RadioContamArOne{} at $148\,\rm GHz$, which is
\RadioContamArOneFrac{} of the typical
decrement signal in our sample.

\section{Peculiar velocities}
\label{s-vpec}

In addition to its Hubble-flow velocity, a galaxy cluster 
can have a velocity offset with respect to its local CMB 
rest-frame, known as its peculiar velocity $v_p$.
As well as being interesting in its own right
as a statistical cosmological probe of large-scale
structure \cite[e.g., ][]{dore03}, $v_p$
introduces a frequency-independent temperature fluctuation 
known as the kinetic SZ (kSZ; Equation \ref{e-kSZ}), in contrast to
the frequency-dependent SZ signal due to thermal motion of the electrons
in the cluster (tSZ; Equation \ref{e-tSZ}).
This kSZ contribution can therefore affect the scaling 
between thermal $Y_{\rm SZ}$ and cluster mass.

\citet{mroc12} report that their $268\,\rm GHz$
data for the triple-merger cluster MACS\,J0717.5$+$3745 
adequately fit their SZE model only if they allow for 
a strong kSZ distortion from a high-velocity subcomponent.
It is unclear whether strong kSZ distortions like those 
seen by \citet{mroc12} are common in merging clusters.
Recent results from  
\citet{plan13xiii} only constrain the RMS fluctuations
in $v_p$ to be $\leq 800\,\rm km\,s^{-1}$ (95\% confidence) 
for a sample from the Meta Catalogue X-ray Detected Clusters
\citep{piffaretti2011}.  If transient kSZ distortions are common
in merging cluster systems, then the
$Y_{\rm SZ}$--mass relation in SZE clusters may be
significantly affected.
As a cautionary indicator here, \citet{sifo13} 
find at least one 
indication of dynamical activity in 
$81^{+19}_{-22}\%$ (13/16) of a representative 
sample of ACT clusters (nine of which overlap 
with this paper's sample).

We use our point source-subtracted multi-band
data to constrain the peculiar velocities of
our targets by parametrizing the observed
SZE signal with $v_p$ and an integrated SZE signal
$Y^{\prime}_{\rm SZ}$.
The iterative pipeline 
(Section \ref{sss-laboca-reduction}) faithfully 
recovers extended emission only up to angular 
scales of $\sim 2^{\prime}$, while $\theta_{500c}$
values can be much larger.  Therefore, to
strengthen constraints while fitting for $v_p$, 
we choose circular apertures for each cluster
(typically 1.5--$2^{\prime}$ radii) that contain
the observed scale of emission in the $345\,\rm GHz$ 
images.  
$Y^{\prime}_{\rm SZ}$ is therefore the effective integrated
SZ signal within each of these apertures, and a lower limit
on $Y_{\rm SZ}(\theta\leq \theta_{500c})$.
We have recently obtained {\em Chandra} imaging
(PI: Hughes) for many clusters in our sample and  
use these data to independently measure $T_e$ in these 
systems (Hughes et al. in prep).
The systematic errors
due to the different weighting between the 
X-ray-derived electron temperatures ($\propto n_e^2$)
and the optical depth-weighted electron temperatures used in
the calculations of the SZ effect are expected to be
only $10$--$20\,\rm km\,s^{-1}$ \citep{sehgal2005}.
For the current analysis, we set 
$k_{\rm B}T_e=5\,\rm keV$ (see Table \ref{t-sze-signal}) for
clusters with unknown gas temperatures.  

For each integrated flux density measurement $S_{\rm SZ,i}$
at $x_i=h\,\nu_i/k_{\rm B}T_{e}$ we compute the expected
SZE signal 
$S^{\rm model}_{\rm SZ}(Y_{\rm SZ}^{\prime},x_i,v_p,T_{e})$.
We fit the data to the multi-band model
using a grid-based search.  The noise in all images 
is nearly Gaussian, and therefore the
conditional probability density  
of obtaining measurements $S_{{\rm SZ},i}$, given the 
model parameters 
$\theta=(Y^{\prime}_{\rm SZ},v_{\rm pec})$ is
\begin{equation}
p_i(x_i;\theta)\propto e^{-(S_{{\rm SZ},i}-S_{{\rm SZ},i}^{\rm model})^2/2\sigma_i^2}.
\end{equation}  The joint probability density of
obtaining all measurements is then given by
$P(x;\theta)\propto \Pi_i p_i$.
We take as the best-fit parameters those that maximize
the likelihood function, and the final quoted
$1\sigma$ uncertainties
in $v_p$ and $Y_{\rm SZ}^{\prime}$ are found by
marginalizing over the 2D likelihood parameter space
and integrating the resulting 1D probability 
functions within iso-contours to contain 
$\pm 68\%$ probability.  
The best-fit peculiar velocities
and corresponding $68.2\%$ confidence intervals
are presenting in Table \ref{t-sze-signal}.

\begin{figure*}
\centering
\includegraphics[scale=1]{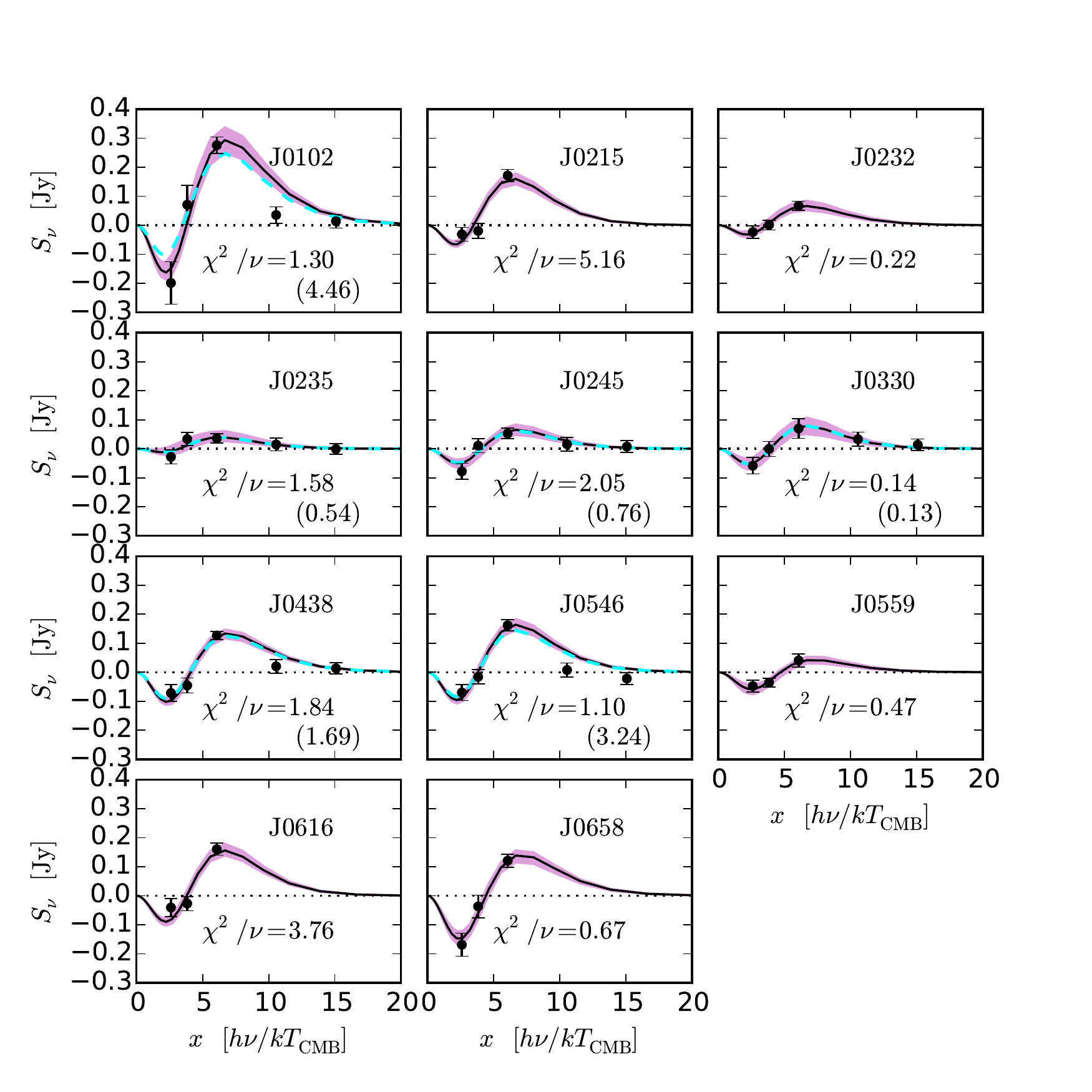}
\caption[Best-fit SZE spectra]
{Best-fit SZE spectra for all 11 clusters in our sample.
The solid black curve and purple region show the
best-fit model and 68\% 
confidence region for 148, 218, and $345\,\rm GHz$ photometry
and the dashed red line shows the best-fit model when including
the {\em Herschel} SPIRE bands at 250, 350, and $500\,\rm \mu m$.
The fit quality ($\chi^2/\nu$) when using only ACT + LABOCA bands 
is indicated in the panels, with the corresponding value when
including SPIRE photometry shown in parentheses.
From left to right, the data points are
ACT\,$148\,\rm GHz$, ACT$\,218\,\rm GHz$, LABOCA$\,\rm 345\,\rm GHz$, SPIRE\,$500\,\rm \mu m$ 
and SPIRE\,$350\,\rm \mu m$ (the latter two only for the six clusters with 
$Herschel$ observations) .}
\label{f-sze-panels}
\end{figure*}

At 148 and 345\,GHz, the SZE signal is 
much brighter than the background confusion
noise.  The situation is reversed for 
350 and $500\,\rm \mu m$, where great efforts must
be made to extract the SZE signal from the strong
SMG background (i.e., see Section \ref{ss-spire_pipeline}).
Recent work by \citet{zemc13} has shown that by subtracting
bright submillimeter point sources near massive clusters,
one will produce a deficit in the surface brightness of the
local cosmic infrared background compared to an 
off-cluster sky region.  By subtracting point-sources 
at a level comparable to that used in Section 
\ref{ss-spire_pipeline}, \citet{zemc13} find a typical
intensity deficit of $\sim 0.5\,\rm MJy\,sr^{-1}$ in the cores
of four massive clusters.  This may related to why
our SPIRE $500\,\mu\rm m$ photometry is systematically
low compared to the other bands (see Figure \ref{f-sze-panels}).
Because of this extra uncertainty in the SPIRE photometry, 
we present peculiar velocity fits for individual clusters 
both including and excluding the SPIRE  
data points (Table \ref{t-sze-signal}); other sample-wide
properties are computed with only ACT + LABOCA photometry.
Figure \ref{f-sze-panels} shows
the best-fit SZE spectra for all
clusters, with the $\chi^2$ per degree of freedom
($\chi^2/\nu$) and $p$ values
indicated in the panels.

The mean peculiar velocity of the sample is \MeanVpec{}, 
consistent with the limit of
 $72\pm 60\,\rm km\,s^{-1}$
found by \citet{plan13xiii}
using the variance of the CMB 
towards X-ray detected clusters.
The measured peculiar velocity dispersion 
of the clusters in our sample is
\vpecVariance{}, and represents 
an upper limit to the intrinsic peculiar velocity dispersion.
\citet{halv09} find that the mass-weighted SZE-derived
temperature measured for the Bullet cluster 
\citep[$10.8\pm 0.9\,\rm keV$; ][]{halv09} is
significantly lower than that derived from
the X-ray observations \citep[$13.9\,\rm keV$; ][]{govoni2004}.
This difference is caused by the different sampling
functions for X-ray derived temperatures ($\propto \int n^2$)
and SZE-derived temperatures ($\propto \int n$).
We adopt the SZE-based temperature for the Bullet cluster from
\citet{halv09} and derive a large peculiar velocity 
of $3100^{+1900}_{-1500}\,\rm km\,s^{-1}$.  
This signal is unlikely to
be caused by contamination of sources fainter than
$S_{345}<10\,\rm mJy$, because positive signal contamination
acts to push $v_{\rm pec}$ to more negative values.
In the Bullet cluster, \citet{markevitch2006} find a 
central ``bullet'' collision velocity of $4700\,\rm km\,s^{-1}$.
Our signal may be due to a kSZ distortion related
to this high-velocity subcomponent, similar 
to the findings of \citet{mroc12} for 
MACS\,J0717.5$+$3745.

\subsection{Impact of $v_{\rm pec}$ on $Y_{SZ}$-$M$ scaling relations}

We next compute $Y^{\prime 0}_{\rm SZ}$, the best-fit
integrated SZ signal when the peculiar velocity is fixed at
$v_{\rm pec} = 0\,\rm km\,s^{-1}$, and define the
systematic uncertainty due to non-zero $v_{\rm pec}$
as $\Delta Y_{\rm SZ}^{\prime} = \left< \left| Y_{\rm SZ}^{\prime} - Y_{\rm SZ}^{\prime 0}\right|\right>$.
We use the $Y_{\rm SZ}-M$ scaling relation 
$M\propto Y_{sz}^{0.48\pm 0.11}$ from \citet{sifo13} to 
derive a relative systematic uncertainty in cluster 
mass due to the clusters' peculiar 
velocities of \fracMassError{}.  Because the
fractional change in $\Delta I_{\rm SZ}$ due to a non-zero 
peculiar velocity is independent of the mean
Compton parameter, our aperture-based estimate of the
relative uncertainty in $Y_{\rm SZ}$ can be directly
applied to cluster-integrated $Y_{\rm SZ}$ measurements.

\subsection{Comparison to simulations}

To understand the  intrinsic scatter and 
systematic uncertainties in our derived values for 
$v_{\rm pec}$, we applied our technique of extracting 
$v_{\rm pec}$ (Section \ref{s-vpec}) to 
the realistic simulated cluster images from the all-sky
maps of \citet{sehgal2010}.  We use the 
``Full SZ''\footnote{http://lambda.gsfc.nasa.gov/toolbox/tb\_sim\_ov.cfm} 
images at 148, 219, and $350\,\rm GHz$, which contain thermal and 
kinetic SZ signals from clusters, groups, and the intracluster medium.
The large scale mass distribution is created using a dark matter
N-body simulation with a comoving box size of $L=1000\,h^{-1}\,\rm Mpc$ 
\citep[for details, see ][]{sehgal2007,sehgal2010}.
We cut out $0.3^{\circ}\times 0.3^{\circ}$ images around 100 
simulated clusters that were randomly chosen among those with
$z>0.3$ and $M_{\rm vir} > 3\times 10^{14}\,M_{\odot}$. 
For each cluster image, we applied the
Fourier-based LABOCA filtering, extracted the
total flux densities at 148, 219, and 350GHz within a $1.5^{\prime}$ aperture
centered on the central halo position, and performed
the maximum likelihood fit from Section \ref{s-vpec}.

We find an intrinsic scatter in output $v_{\rm pec}$
of $\sigma_{v_{\rm pec}} = 213\,\rm km\,s^{-1}$, and a bias
of ($\left< v_{\rm pec, out}/v_{\rm pec, in}\right> - 1$) of 0.36.  
For $0.25^{\prime}$ apertures, the scatter is 
$137\,\rm km\,s^{-1}$ and the median bias is 0.15.  
Given that the smaller apertures reduce the bias 
and scatter, we attribute both effects to
the aperture-dependent irreducible error from assuming 
a single temperature
and velocity throughout the volume of a cluster.

\begin{figure}
    \centering
    \begin{tabular}{c}
           \includegraphics[scale=0.65]{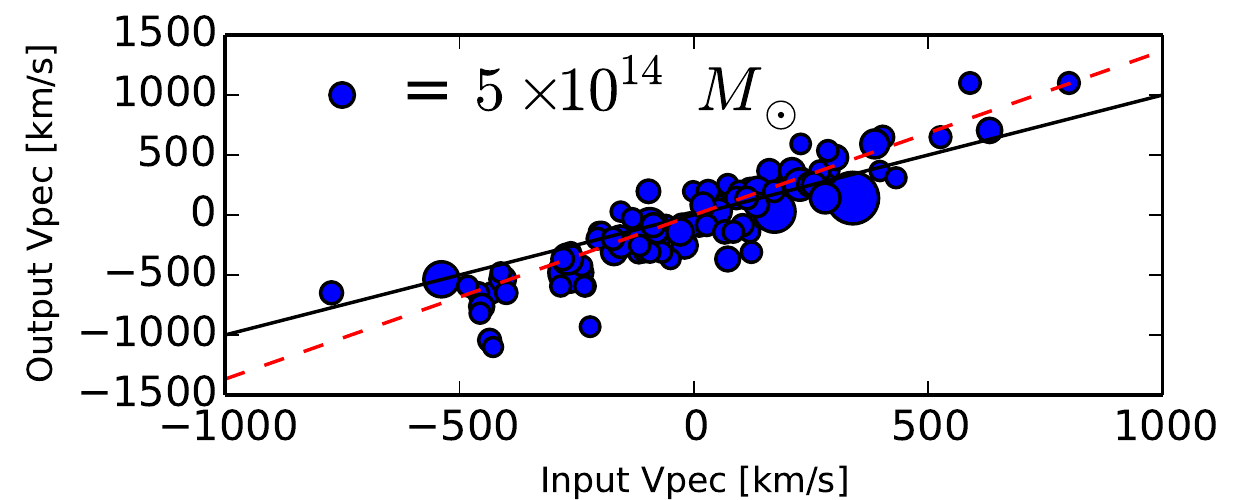} \\
        \includegraphics[scale=0.65]{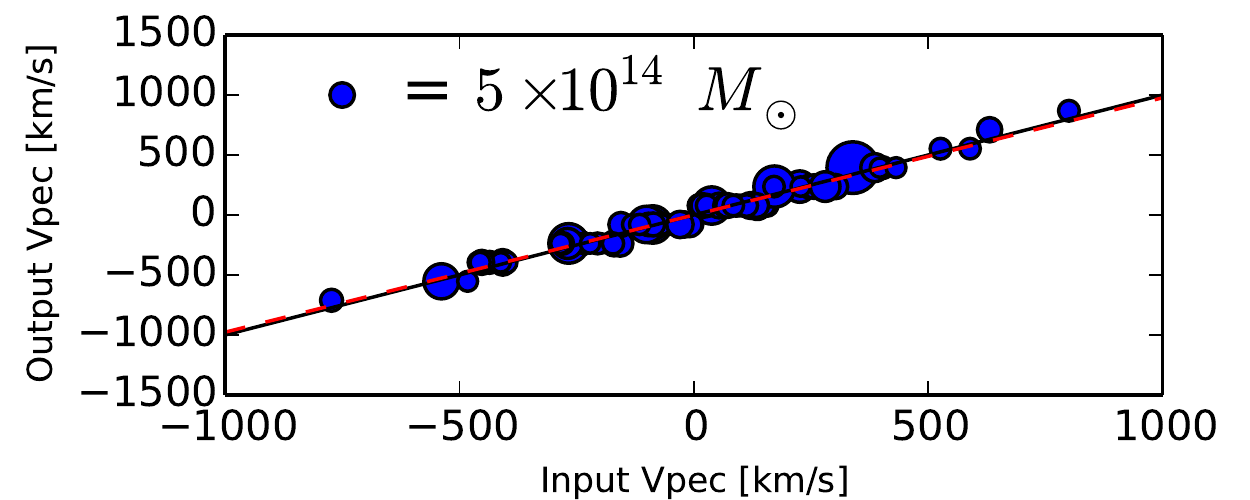} \\
        \includegraphics[scale=0.65]{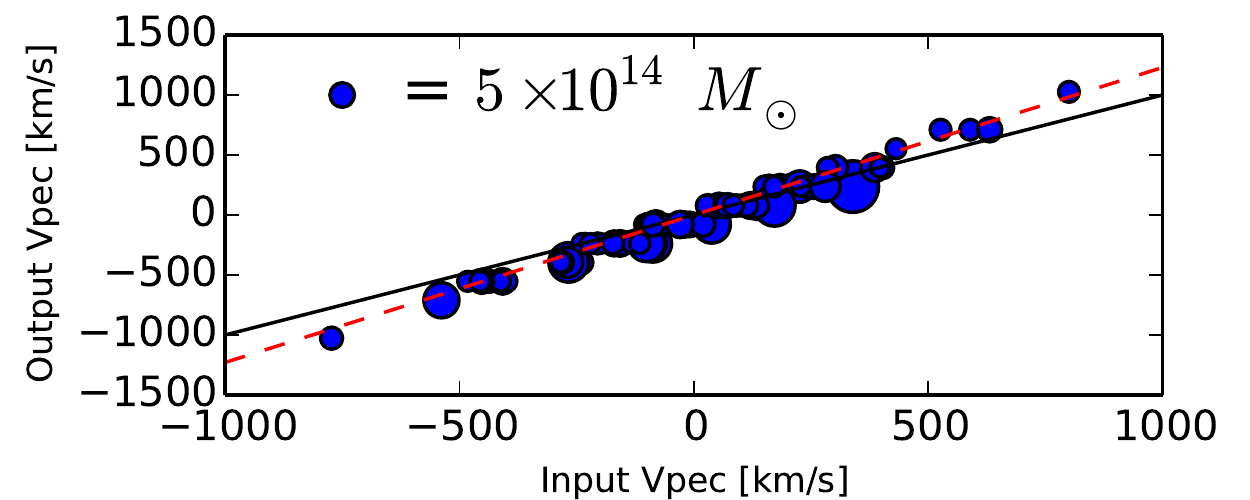} 
    \end{tabular}
    \caption{Recovered (output) peculiar velocity as a function of
             known (input) peculiar velocity for a sample of
             100 clusters with $z>0.3$ and  $M_{\rm vir} > 3\times 10^{14}\,M_{\odot}$
             from the \citet{sehgal2010} all-sky simulations of the microwave
             sky.  {\em Top panel:}  Photometry is realistically extracted
             from the simulated images using $1.5^{\prime}$-radius circular 
             apertures and $T_{\rm fit}=T_{\rm true}$.
             {\em Middle panel:} Perfect photometry with 
             $T_{\rm fit}=T_{\rm true}$.  {\em Bottom panel: }
             Perfect photometry with $0.8\times T_{\rm fit} =T_{\rm true}$.
             In each panel, the solid black line shows a linear relation
             with unity slope, and the dashed red line shows the best 
             least-squares fit of a line constrained to pass through the
             origin.  The diameters of the symbols are proportional to the
             simulated cluster masses with the scale of $5\times 10^{14}\,M_{\odot}$
             indicated at the upper left of each panel.}
    \label{f-sims}
\end{figure}

To investigate the source of the scatter and bias in 
more detail, we next performed an experiment exactly the 
same as above, except that we {\em computed} the photometry 
for each system while varying the relation between 
the gas temperature assumed in the maximum likelihood fit
$T_{\rm fit}$, and the true gas temperature of the cluster 
$T_{\rm true}$.  When $T_{\rm fit}=T_{\rm true}$, we find 
negligible scatter and bias, but for 
$T_{true} = 0.8\,\times T_{fit} $,
we find a median bias in the recovered peculiar velocity of
0.2.  This induced temperature-related bias can explain the 
bias we see in the extracted $v_{\rm pec}$ of the simulated clusters 
above: in the simulations, $T_{\rm true}$
is measured in the 3D centers of the clusters, and larger 
apertures will include more cool gas at larger cluster-centric 
radii.   This effect is also
likely operating in the real observations since the X-
ray-derived gas temperatures are weighted by $n_e^2$
and preferentially sample the dense interiors of clusters.
This might explain, for example, the $22\%$ difference found
by \citet{halv09} in SZE- and X-ray-derived temperatures
quoted above.
Figure \ref{f-sims} shows the input and output peculiar velocities
for each of the three numerical experiments above.

\section{Comparison to previous work}
\label{s-discuss}

\begin{figure}
    \centering
   \includegraphics[scale=0.6]{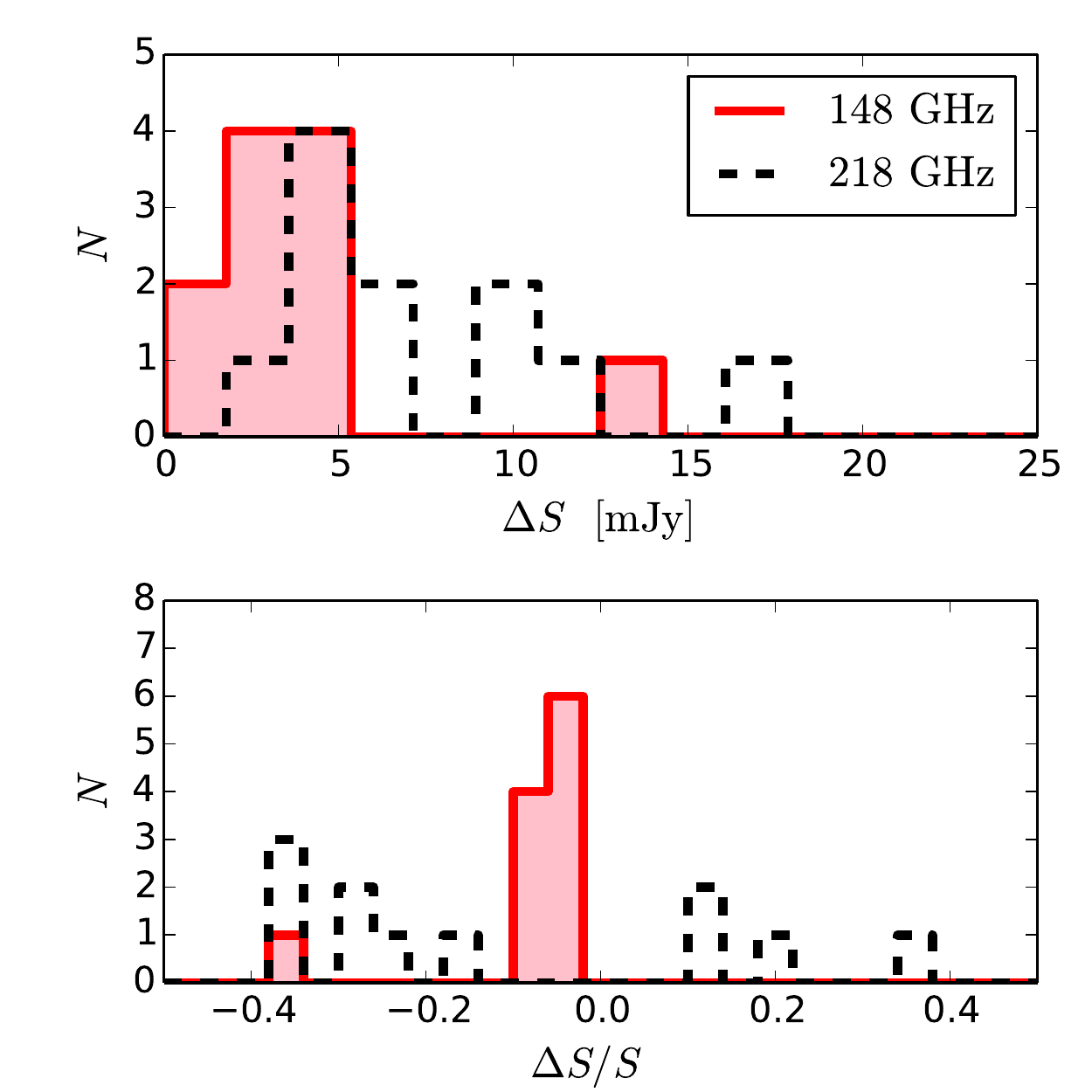}
   \caption{148 and 218 GHz absolute ({\em above}) 
   and fractional ({\em below}) contamination of
   the SZ signal by SMGs and radio point sources. 
   The varying sign of the $218\,\rm GHz$ fractional contamination
   ($\Delta S / S$) is caused by noise fluctuations
   because the clusters typically have small or negligible 
   $218\,\rm GHz$ signals.
   }
    \label{f-histos}
\end{figure}
Figure \ref{f-histos} shows the combined 
absolute and relative contamination levels from
radio sources and SMGs in our sample.
Our results are in general agreement with previous
predictions and measurements of
radio and infrared source contamination at $148\,\rm GHz$.
For example, based on simulations of the microwave sky, 
\citet{sehgal2010} predicted that $<20\%$ of clusters 
with $M_{200c} > 10^{14}\,M_{\odot}$ and $z < 0.8$
should have their decrements filled in by $>50\%$.
In our sample of clusters, we find a mean individual
contamination fraction of 
$\left< \Delta S / S \right> = 8\%$, and none 
with contamination $>50\%$.

\citet{lima2010} presented analytic predictions 
of the Poisson variance in SZ flux density due to
lensed background SMG populations.  At 148\,GHz,
\citet{lima2010} predict $\sigma_{\rm lensed\,SMG}\sim 2$--$3\,\rm mJy$
within $r_{2500c}$.  Within our 1--$2^{\prime}$ radius apertures
(similar to $r_{2500c}$ for our cluster sample), we 
estimate the RMS intensity of $148\,\rm GHz$  fluctuations due
only to the significantly detected LABOCA point sources
to be $2\,\rm mJy$.

Using 1.4, 30, and $140\,\rm GHz$ observations of 45 massive
clusters, \citet{sayers2013b} estimate the fractional contamination
due to radio sources to be $\simeq 20\%$ on average, with only 1/4
of clusters showing contamination greater than $1\%$.  The mean
individual contamination ($\left<\Delta S/ S \right>$) by radio 
sources in our sample is 6\%.

\citet{bens03} constrained the peculiar velocities
of six clusters using the Sunyaev-Zel'dovich
Infrared Experiment in three frequency bands
between 150 and 350GHz, and placed a constraint on the intermediate
Universe bulk flow velocity of $<1420\,\rm km\,s^{-1}$ 
(95\% confidence), consistent with ours.

\section{Conclusions}
\label{s-conclusions}

We present high-resolution $345\,\rm GHz$
LABOCA, $2.1\,\rm GHz$ ATCA (10 of 11) 
and $Herschel$ SPIRE (6 of 11) 
imaging of eleven massive SZE-selected clusters from the ACT 
southern survey \citep{marr11,mena10}.  We use
these data to constrain the
levels of radio source 
and SMG contamination of the SZE signals of the 
clusters, and also to constrain the cluster
peculiar velocities using the kSZ effect.

We find that the contamination by $2.1\,\rm GHz$
radio sources of the $148\,\rm GHz$ SZE decrement 
is \RadioContamArOne{} per cluster, or
$\sim$\RadioContamArOneFrac{} of the 
SZE decrement signal in our analysis.  
We measure the $2.1\,\rm GHz$ number counts
in three disjoint regions around the clusters, 
$\theta < \theta_{2500c}$, 
$\theta_{2500c} < \theta < \theta_{500c}$,
and $\theta_{500c} < \theta < 5\,\theta_{500c}$,
and find an enhancement in the counts for 
$\theta < \theta_{2500c}$.

The typical contamination from bright, unresolved SMGs is 
\SmgContamLaboca{} (\SmgContamLabocaFrac{}) per 
cluster at $345\,\rm GHz$, scaling to a
\SmgContamArOneFrac{} effect on the 
$148\,\rm GHz$ decrement.  The number counts of
bright $S_{345}>10\,\rm mJy$ SMGs are $\sim 10\times$
greater than blank field measurements, likely due to
gravitational lensing by the clusters' 
potentials \citep[as seen in, e.g., ][]{knud08,joha11}.
The combined contamination by SMGs and radio sources ($\sim 5\%$ of the
148\,GHz decrement signal on average) 
may contribute to, but still remains less than, the scatter found in 
the $Y_{\rm SZ}$-to-mass scaling relation of SZE clusters.

After subtraction of the bright SMGs, we
use our multi-band data to constrain the peculiar 
velocities of the clusters.
For clusters with high-significance SZE detections, 
the typical uncertainty in 
$v_{\rm pec}$ is $\pm 1000\,\rm km\,s^{-1}$, and
for the full sample we find a mean peculiar 
velocity of \MeanVpec{}.
By comparing the best-fit $Y^{\prime}_{\rm SZ}$
values with and without fitting for $v_{\rm pec}$,
we estimate that peculiar velocities introduce a scatter
to the SZ-estimated mass of clusters at the level
of \fracMassError{}.  

Future observations with
higher angular resolution and support for multiple
instantaneous millimeter/submillimeter bandpasses 
can help reduce the uncertainties on $v_{\rm pec}$.
Higher angular resolution can resolve a larger 
fraction of the CIB into point sources, and therefore 
reduce the magnitude of the remaining confused SMG 
background. The capability to observe multiple
wavebands spanning the SZE decrement to
the increment would allow the wavebands to
be analyzed homogeneously, reducing the
complexity of data reduction and related
systematic uncertainties.  These capabilities
can be provided by wide-format multi-band
bolometer cameras on telescopes like
the Large Millimeter Telescope and the
upcoming Cerro Chajnantor Atacama 
Telescope.

\section{Acknowledgments}
\acknowledgments
RRL and AJB acknowledge significant support for this work from the U.S. National 
Science Foundation through grant AST-0955810.  JPH acknowledges support
from the National Aeronautics and Space Administration (NASA) through Chandra 
Awards numbered GO1-12008X and GO2-13156X 
issued to Rutgers University by the Chandra X-ray 
Observatory Center, which is operated by the Smithsonian Astrophysical 
Observatory for and on behalf of NASA under contract NAS8-03060,
and through an award issued by JPL/Caltech in association with {\it Herschel},
which is a European Space Agency Cornerstone Mission with significant 
participation by NASA.

ACT operates in the Parque Astron\'omico Atacama in northern Chile under 
the auspices of the Programa de Astronom\'ia de la Comisi\'on Nacional de 
Investigaci\'on Cient\'ifica y Tecnol\'ogica de Chile (CONICYT). This work 
was supported by the U.S. National Science Foundation through awards
AST-0408698 and AST-0965625 for the ACT project, and PHY-0855887, 
PHY-1214379, AST-0707731, and PIRE-0507768 (award No. OISE-0530095). 
Funding was also provided by Princeton University, the University of 
Pennsylvania, and a Canada Foundation for Innovation (CFI) award to UBC. 
Computations were performed on the GPC supercomputer at the SciNet HPC 
Consortium. SciNet is funded by the CFI under the auspices of Compute 
Canada, the Government of Ontario, the Ontario Research Fund --- Research 
Excellence, and the University of Toronto.
We acknowledge support from the FONDAP Center for Astrophysics 15010003,
BASAL CATA Center for Astrophysics and Associated Technologies.

The authors thank the APEX staff for their help in carrying out the 
observations presented here, as well as Phil Edwards, Robin Wark, and
Shane O'Sullivan for their assistance with the ATCA observations.
We thank the anonymous referee for helpful feedback that has
improved this manuscript.
APEX is operated by the 
Max-Planck-Institut f\"ur Radioastronomie, the European Southern 
Observatory, and the Onsala Space Observatory.

\end{document}